\documentclass[fleqn,usenatbib]{mnras}
\usepackage{newtxtext,newtxmath}
\usepackage{lipsum}
\usepackage[T1]{fontenc}
\usepackage{tgbonum}

\DeclareRobustCommand{\VAN}[3]{#2}
\let\VANthebibliography\thebibliography
\def\thebibliography{\DeclareRobustCommand{\VAN}[3]{##3}\VANthebibliography}

\usepackage{graphicx}	
\usepackage{amsmath}	
\usepackage{float}
\usepackage{here}
\usepackage{cancel}
\usepackage{natbib}

\DeclareMathOperator\erf{erf}

\newcommand{\HI}{H{\textsc{i}}}

\title[HI-catalogue  and source finding methods]{MIGHTEE-HI: \HI\ catalogue of 293 sources for the COSMOS field and comparative study of 3-dimensional source finding methods}

\author[M. Maksymowicz-Maciata et al.]{
Michalina Maksymowicz-Maciata,$^{1}$\thanks{E-mail: michalina.maksymowicz-maciata@bristol.ac.uk}
Natasha Maddox,$^{1}$
Catherine Hale,$^{2,5}$
Ben Maughan,$^{1}$
Matt J. Jarvis,$^{2}$
\and
Anastasia A. Ponomareva,$^{3,2}$
Ian Heywood,$^{2,4,5}$
Hengxing Pan,$^{6,7}$
Sushma Kurapati,$^{8}$
Tom G. Hardy,$^{9}$
\and
Marcin Glowacki,$^{10,11}$
Tobias Westmeier,$^{12}$
Maarten Baes,$^{13}$
Seoyoung Lyla Jung,$^{2}$
Andreea A. V\u ar\u a\c steanu$^{2}$
\\
$^{1}$School of Physics, H.H. Wills Physics Laboratory, Tyndall Avenue, University of Bristol, Bristol, BS8 1TL, United Kingdom\\
$^{2}$Sub-Dep. of Astrophysics, Dep. of Physics, University of Oxford, Denys Wilkinson Building, Keble Road, Oxford OX1 3RH, United Kingdom\\
$^{3}$Centre for Astrophysics Research, School of Physics, Astronomy and Mathematics, University of Hertfordshire, College Lane,
Hatfield, AL10 9AB, UK \\
$^{4}$SKA Observatory, Jodrell Bank, Lower Withington, Macclesfield, SK11 9FT, UK \\
$^{5}$Department of Physics and Electronics, Rhodes University, PO Box 94, Makhanda, 6140, South Africa \\
$^{6}$National Astronomical Observatories, Chinese Academy of Sciences, Beijing 100101, People’s Republic of China\\
$^{7}$Guizhou Radio Astronomical Observatory, Guizhou University, Guiyang 550000, China \\
$^{8}$Netherlands Institute for Radio Astronomy (ASTRON), Oude Hoogeveensedijk 4, 7991 PD Dwingeloo, The Netherlands\\
$^{9}$Department of Physics, Durham University, South Road, Durham DH1 3LE, UK \\
$^{10}$Institute for Astronomy, University of Edinburgh, Royal Observatory, Edinburgh, EH9 3HJ, United Kingdom\\
$^{11}$Inter-University Institute for Data Intensive Astronomy, Department of Astronomy, University of Cape Town, Cape Town, South Africa\\
$^{12}$International Centre for Radio Astronomy Research (ICRAR), The University of Western Australia,
35 Stirling Highway, Crawley WA 6009, Australia\\
$^{13}$Sterrenkundig Observatorium, Universiteit Gent, Krijgslaan 299, B-9000 Gent, Belgium\\
}

\date{Accepted XXX. Received YYY; in original form ZZZ}

\pubyear{\the\year{}}

\begin{document}
\label{firstpage}
\pagerange{\pageref{firstpage}--\pageref{lastpage}}
\maketitle

\begin{abstract}
We present a catalogue of \HI\ sources extracted from the MIGHTEE survey data cubes covering the COSMOS field. The catalogue contains 293 sources in the redshift range of $0.004<z<0.093$. In addition to \HI\ masses and velocity widths, the catalogue includes optical through near-infrared photometry and inferred stellar masses and star-formation rates. The quantity of sources in the \HI\ catalogue acquired through untargeted source finding is greatly influenced by the source finding methods used. This study therefore also provides a well-characterised expected completeness of the detected sample of galaxies based on their properties, informing of any detection biases, inferred through a comparative study of different source finding algorithms. We have tested the performance of widely-used source finders: PyBDSF, ProFound and SoFiA, along with new source finder LESHI, focusing exclusively on \HI\ source detection rather than source characterisation in the first instance. The source finders were tested by injecting a sample of simulated galaxies divided into narrow bins of mass, inclination and distance into a MeerKAT data cube. The results inform the source finding strategies for the MeerKAT International GigaHertz Tiered Extragalactic Exploration (MIGHTEE) survey, as well as upcoming SKAO surveys. 

\end{abstract}




\section{Introduction}

\begin{figure*}
    \centering
    \includegraphics[width=0.85\linewidth]{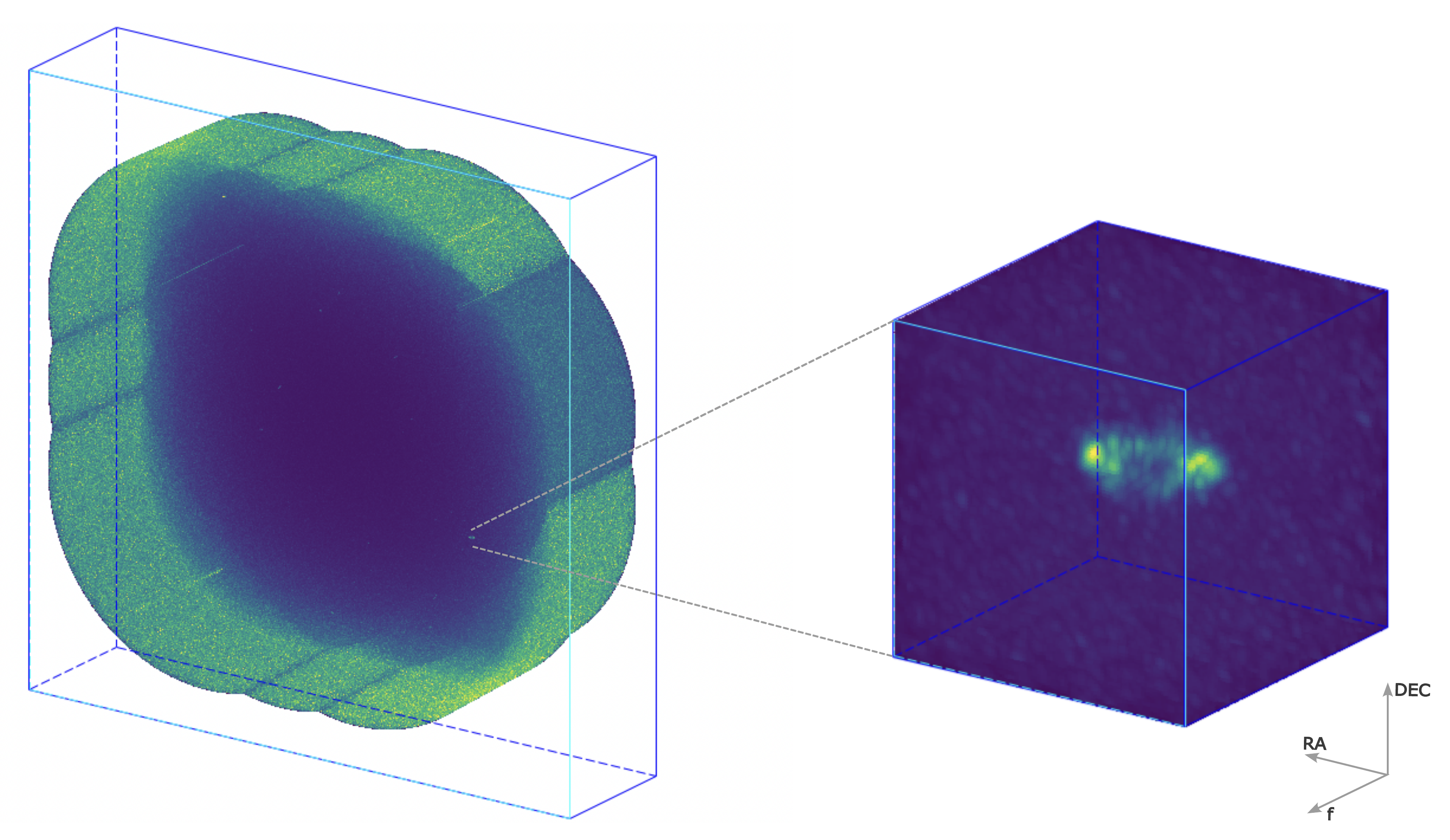}
    \caption{Three-dimensional view of a data cube for the frequency range of 1.3685-1.3961 GHz (spanning 1055 channels) used for source injection in this work (left) and an extracted small volume  centred on an example of a real source (right), both adapted from SAOImageDS9 application \citep{2019zndo...2530958J} visualization. The data cube extract covers the COSMOS field with a mosaic of 15 pointings with a total area spanning $\sim4$ deg$^2$ (4600 by 4600 pixels), centred on RA$=150.03158$ deg, DEC$=2.208856$ deg, taken by the MeerKAT telescope as part of the MIGHTEE survey.}
\label{fig1}
\end{figure*}

The 21 cm spectral line emission of \HI\ is one of the primary wavelengths observed in radio astronomy as neutral hydrogen serves as the raw material for the build-up of stellar mass. It can be observed in galaxies, tracing their structure and neutral gas reservoir, and free-floating clouds that have been stripped from galaxies, and is of great scientific interest (see for example \citealt{2001AJ....122.2969H}, \citealt{2021MNRAS.506.2753R} and references therein). The \HI\ emission is intrinsically very faint (compared to optical data) and mapping the extragalactic \HI\ universe therefore requires wide, deep and untargeted surveys, leading to a large number of sparsely-populated data cubes.

Several large-area \HI\ surveys have already been undertaken. Among single dish radio telescope surveys is the Arecibo Legacy Fast ALFA Survey (ALFALFA; \citealt{2005AJ....130.2598G}), which uses the observations from the Arecibo Telescope. The whole sky is covered by the southern hemisphere \HI\ Parkes All-Sky Survey (HIPASS; \citealt{2001MNRAS.322..486B}), along with the Northern HIPASS extension (NHICAT; \citealt{2006MNRAS.371.1855W}). Other large-sky surveys include the Commensal Radio Astronomy FAST Survey (CRAFTS; \citealt{2018IMMag..19..112L}), and the FAST All Sky \HI\ survey (FASHI; \citealt{2024ApJ...971..131Z}), which utilise the Five hundred meter Aperture Spherical Telescope (FAST; \citealt{2011IJMPD..20..989N}). Interferometric and wide surveys include \HI\ surveys with Apertif (\citealt{2022A&A...667A..38A}), which is a phased-array feed system for the Westerbork Synthesis Radio Telescope, the Widefield ASKAP L-band Legacy All-sky Blind surveY (WALLABY; \citealt{2020Ap&SS.365..118K}) and the Deep Investigation of Neutral Gas Origins (DINGO; \citealt{2023MNRAS.518.4646R}), which use the Australian Square Kilometer Array Pathfinder (ASKAP; \citealt{2021PASA...38....9H}). Among deeper interferometry surveys is the COSMOS \HI\ Large Extragalactic Survey (CHILES; \citealt{2015AAS...22542703F}) targeting higher-redshift sources with the upgraded Karl G. Jansky Very Large Array (VLA; \citealt{2020PASP..132c5001L}), and the Blind Ultra Deep HI Environmental Survey (BUDHIES; \citealt{2013MNRAS.431.2111J}), using the the Westerbork Synthesis Radio Telescope (WSRT; \citealt{1974A&A....33..289H}). The Karoo Array Telescope (MeerKAT; \citealt{2009IEEEP..97.1522J}, \citealt{2016mks..confE...1J}) is used to study \HI\ by the Looking At the Distant Universe with the MeerKAT Array survey (LADUMA; \citealt{2012IAUS..284..496H}, \citealt{2016mks..confE...4B}), and the MeerKAT International GigaHertz Tiered Extragalactic Exploration survey (MIGHTEE; \citealt{2016mks..confE...6J}), which we further describe in Section \ref{MIGHTEE}. The continuing operation of existing radio telescopes, as well as the upgrade of their resolution and field of view and construction of new telescopes, such as the Square Kilometre Array (SKA; \citealt{2015aska.confE.174B}), DSA-2000 radio camera (\citealt{2019BAAS...51g.255H}), or the Canadian Hydrogen Observatory and Radio-transient Detector (CHORD; \citealt{2019clrp.2020...28V}), will lead to a rapid increase in the number and size of data cubes and their sensitivity. The MeerKAT radio telescope alone can produce 4.7 GB of data per second.

It is therefore a challenge of great importance to find the best approach to optimally and reliably perform source finding. This is particularly challenging for three-dimensional spectral line data (data cubes), which, with the new generation of radio-telescopes, can span a wide range of channels and pixels (e.g. Fig. \ref{fig1}), making the search for the signal much more computationally expensive due to the added frequency dimension. Moreover, the challenge increases for faint sources close to the noise level, as their inclusion may lead to many false positives, forcing trade-offs between sample completeness (fraction of sources found) and reliability/purity (fraction of found sources that are real).

It has been shown that using eyes remains one of the most complete and reliable methods of source finding (\citealt{2025A&A...696A.113T}), however applying it to large amount of data is unfeasible. Consequently a number of automated source finders emerged from the scientific community. Among two-dimensional imaging source finders designed for optical data is ProFound, which works well for all kinds of images including radio continuum data (\citealt{2018MNRAS.476.3137R}, \citealt{2019MNRAS.487.3971H}; see Section \ref{sec:sofia} for more details). Other two-dimensional source finders are Caesar (\citealt{2019PASA...36...37R}), Python Blob Detector and Source Finder (PyBDSF; \citealt{2015ascl.soft02007M}; see Section \ref{sec:sofia} for more details), Aegean (\citealt{2012ascl.soft12009H}), and BLOBCAT (\citealt{2012MNRAS.425..979H}). Even though these source finders were not designed for data cubes, but for two-dimensional images instead, their capabilities can still be applied to three-dimensional data with some post-processing (as done in this work for PyBDSF and ProFound). Among the source finders that were designed for three-dimensional data cubes (and therefore particularly relevant for \HI\ source finding) is Source Finding Application (SoFiA; \citealt{2021MNRAS.506.3962W}, \citealt{2015MNRAS.448.1922S}; see Section \ref{sec:sofia} for more details), DUCHAMP (\cite{2012MNRAS.421.3242W}) and LIghtweight Source finding Algorithms (LiSA; \citealt{2022A&C....4100631T}).

Source finders employ different algorithms, therefore it is essential to match the most suitable tool to the data and scientific requirements. For example, measurements of the \HI\ mass function require reliable source identification with well known completeness (defined by the ratio of the number of sources found and the number of all sources in the data), while searches for faint sources benefit from high completeness at the cost of reliability (defined by the ratio of the number of found sources that are real and the total number of found sources), which might be aided by ancillary data.
Therefore, testing and comparing different tools becomes an important step. Some works have already attempted to tackle this problem:
\cite{2023A&A...670A..55B} carried out a comparative study of SoFiA, MTObjects, and supervised deep
learning algorithm originally designed for medical imaging. Many source finders were also tested during the SKA Science Data Challenge 2 (\citealt{2023MNRAS.523.1967H}) on a simulated data product representing a 2000 h SKA-Mid spectral line observation.

In this paper, we test the SoFiA, ProFound and PyBDSF source finders and introduce a new source finder LESHI (Line Emission Source-Hunting Integrator), developed for \HI\ data. We explore how they perform and compare by measuring completeness and purity (reliability) of the resulting samples when run on a cube with simulated injected \HI\ emission line sources and explore their different strengths and weaknesses. In contrast to the past work done on comparing different source finders, we push the testing boundaries by incorporating different ranges of mass, inclination, and frequencies (distances) of the injected sources and investigate how the source finders perform in each parameter bin. Lastly, we derive a function outputting the expected completeness from input mass, inclination and distance bins based on our results.

In the second part of this work, we put the lessons learned from testing the source finders into practice and produce a catalogue of \HI\ sources acquired through untargeted source finding with the LESHI source finder for the MIGHTEE data cubes available for the COSMOS field (\citealt{2024MNRAS.534...76H}). This field has a wealth of deep multi-wavelength ancillary data, both spectroscopic and photometric, allowing us to optically confirm the detections and determine the stellar contents and star formation rates. Our catalogue of 293 galaxies includes \HI\ properties, along with photometry, stellar masses and star-formation rates (SFRs) of the optical counterparts.

The paper is structured as follows. In Section \ref{sec:methods}, we briefly describe the tested source finders and the injecting methods, while in Section \ref{sec:results} we present the results and discuss the strengths and weaknesses of each source finder. Finally, in Section \ref{sec:catalogue} we present the MIGHTEE COSMOS \HI\ catalogue. Section \ref{sec:conclusions} is reserved for summary and conclusions.

Throughout this paper we assume $\Lambda$CDM cosmology with $H_{0} = 70 \text{ km s}^{-1}\text{Mpc}^{-1}$, $\Omega_{\text{M}}=0.3$ and $\Omega_{\Lambda}=0.7$. The magnitudes are given in the AB magnitude system and are not extinction corrected.

\section{Source Finders Comparison Methods}
\label{sec:methods}

\subsection{Source injection}

We assess the completeness and reliability of the source finders, by injecting artificial sources into a real cube. This allows us to see how each source finder performs with real data, with noise and artefacts, that are otherwise difficult to emulate. The cube used to test the source finders is a frequency slab of a full datacube produced by the South African MeerKAT radio telescope as part of the MIGHTEE survey (see \citealt{2024MNRAS.534...76H} and Section \ref{MIGHTEE} for more details). It covers the COSMOS field with a mosaic of 15 pointings with a total area spanning $\sim4$ deg$^2$ and frequency range of 1.3685-1.3961 GHz with channel width of 26.125 kHz. We do not use the full frequency range of the datacube, as we do not need the full volume for the source injection and we avoid frequency ranges impacted by radio frequency interference (RFI). The testing results should hold for all of the frequencies spanned by MIGHTEE data, as the noise does not change significantly across the investigated distances, as can be seen in Fig. 6 of \cite{2024MNRAS.534...76H}. The median channel root mean square (rms) noise is 70 $\mu$Jy/beam in the middle of the mosaic rising to 480 $\mu$Jy/beam toward the edge of the image. The median synthesised beam is $\sim$15" in diameter. The cube has 4600 by 4600 pixels and 1055 channels (see Fig. \ref{fig1}), therefore providing a sparse volume that can be populated with a large number of galaxies without the risk of source confusion. 

The cube contains real sources (barely visible "specs" in the centre region of the cube in Fig. \ref{fig1}) and continuum artifacts ("lines" along the frequency direction in the cube in Fig. \ref{fig1}). To account for the real sources, we have run each source finder on the data (without injecting any new sources) and investigated each source found. The detections that had obvious emission and were within 8 arcseconds of sources in the Sloan Digital Sky Survey (SDSS) optical catalogue were treated as real. Those were not masked, but instead a catalogue of all obvious real sources was created and their detections (or non-detections) were subsequently ignored.

\begin{figure*}
    \centering
    \includegraphics[width=0.995\linewidth]{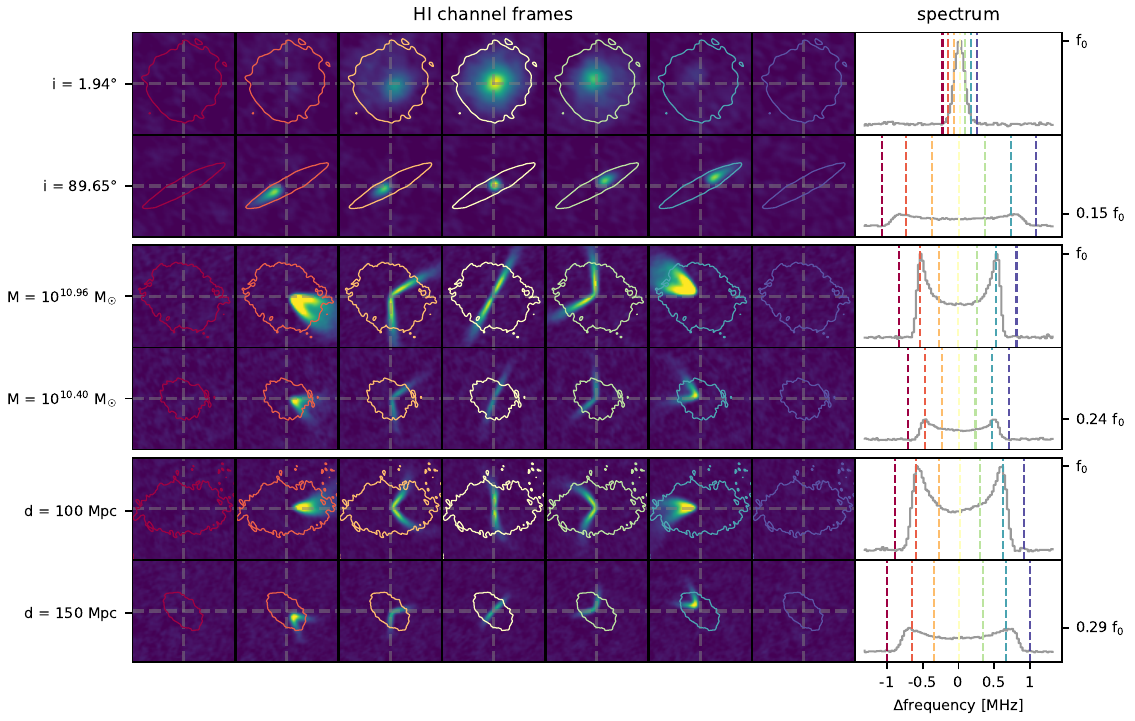}
    \caption{\HI\ emission channel frames and spectra for a sample of simulated galaxies. The top two panels show galaxies of similar mass and distance, but varying inclination, the middle two panels show galaxies of similar inclination and distance, but varying mass, and the bottom two panels show galaxies of similar inclination and mass, but varying distance. Contours enclose the 3$\sigma$ level of the simulated \HI\ emission and are colour-coded (along with the vertical dashed lines on the spectrum plots) by the frequencies for which the channel images are shown. }
\label{fig2}
\end{figure*}

The artificial sources were simulated using the \textsc{$^{\text{3D}}$Barolo} package (\citealt{2015MNRAS.451.3021D}) that uses 3D tilted-ring models of line emission from galaxies. A sample of six simulated galaxies is shown in  Fig. \ref{fig2}. The sources were simulated using the standard \HI\ mass-luminosity conversion (\citealt{2017PASA...34...52M}) and \HI\ mass-size relation (\citealt{2016MNRAS.460.2143W}, \citealt{2022MNRAS.512.2697R}), which governs their physical size given an input HI mass. The rotation curves were derived using the baryonic Tully-Fisher relation (\citealt{1977A&A....54..661T}, \citealt{2021MNRAS.508.1195P}) for dwarf galaxies after assuming the baryonic mass being equal to \HI\ mass and for massive galaxies the rotation velocity was randomly chosen between 100 and 350 km/s. The simulated galaxies were convolved with the synthesised beam of the cube used for injection ($\sim$15"). We injected 600 uniformly-spaced galaxies per run for low mass galaxies ($<10^{8}$M$_{\odot}$) and to avoid overcrowding the cube, 150 per run for high mass galaxies. We divide them into equal groups of different distances and explore the full mass-inclination plane by repeating the runs for different mass and inclination bins of the injected galaxies, resulting in the total of 24000 injected sources. For all of the runs, the cube was at least 99.3\% empty. We note that the cube used for injection spans only the distance range of 75 Mpc - 165 Mpc. However we simulate the galaxies as if they were at luminosity distances of 50, 100, 150 and 350 Mpc (accordingly lowering the flux and projected size) and inject them centred at a random frequency channel within the cube, and so the frequency of the injected galaxy no longer corresponds to the distance. The galaxies were injected in the central region of the data cube, where the noise level can be approximated as uniform. We note however, that higher noise has the same effect as higher distance, consequently lowering the SNR of the emission. We note that the increased noise towards the field edges would need to be accounted for in statistical studies such as measuring the \HI\ mass function (\citealt{2023MNRAS.522.5308P}).

\subsection{Injected source parameters}
The observed signal from a galaxy is greatly influenced by its properties, such as mass, inclination and distance. Figure \ref{fig2} shows how each of these parameters changes the observations in the spatial and frequency dimensions, by plotting the channel frames and spectra for a sample of injected galaxies. As can be seen, the spectrum's height, width and shape change dramatically for different parameters. It is therefore expected that the corresponding inputs for the injected galaxies should dictate how the source finders perform in each of the mass-inclination-distance bins (see Fig. \ref{completeness_histogram}). To investigate the response of the source finders to the parameters of the injected galaxies, we inject the galaxies in narrow bins of mass, inclination and frequency (distance). The ranges and bin-widths are elaborated on in the following subsections.

\subsubsection{Inclinations}
\label{sec:Inclinations}
For randomly oriented galaxies, the distribution of the cosines of the inclination angles should be uniform. Our inclination bins therefore span the range of 0 < cos(i) < 1 in bins of width 0.2, with cos(i) = 0 for edge-on galaxies and cos(i) = 1 for face-on ones. As can be seen in the first panel of Fig. \ref{fig2}, varying the inclination has an impact on how the measured flux is spread across spatial pixels (if the source is resolved) and spectral channels.
For the face-on galaxy (low inclination angle), the width of the spectral line comes mainly from the velocity dispersion of \HI\ clouds, while for the edge-on galaxy (high inclination angle), the width of the emission (and the two peaks) is a result of the galaxy's rotation and the effect of Doppler's shift. While the shape of the spectrum changes significantly, the total flux is conserved, therefore for greater inclinations, while the emission line is wider, the peak flux also must be lower, pushing it towards the noise level, which leads to the galaxy being more challenging to detect. Hence, it is expected that the completeness of source finders should be greater for galaxies of a given mass with lower inclinations (face-on galaxies). However, if the spectral resolution of the data is low enough, emission from face-on galaxies might have the width of only one channel, leading to some source finders to reject it. It is also worth noting that if a high-inclination galaxy is faint enough, the middle section of its double-peaked spectrum can fall below the noise level, which leads to the spectrum looking like two separate peaks, possibly confusing the source finders if their separation is too large to be associated.

\subsubsection{Masses}
The typical \HI\ masses in galaxies span from 10$^{6}$ M$_{\odot}$ to 10$^{11}$ M$_{\odot}$ (\citealt{2016MNRAS.460.2143W}). We have therefore adopted the same range for the mass bins starting at 10$^{6}$ M$_{\odot}$ and incrementing by 1 in logarithmic space until 10$^{11}$ M$_{\odot}$, in agreement with previous MIGHTEE-\HI\ masses at z $<$ 0.1. We increase the resolution down to 0.5 and 0.25 dex, for masses where the completeness is expected to change most rapidly. We increase the minimal mass tested to 10$^{8}$ M$_{\odot}$ at the distance of 350 Mpc, as lower masses would be below the detection threshold at this distance. As can be seen in the second panel of Fig. \ref{fig2}, \HI\ mass has the expected impact on the form of the observed signal. For higher masses, the peak flux is greater and the emission line is slightly wider. The general shape of it remains similar to that from lower mass galaxies, as it is mostly dictated by the inclination.

\subsubsection{Distances} 

The influence of the distance to the galaxy is very straightforward. Since measured flux is inversely proportional to the luminosity distance squared, the galaxy will simply appear dimmer (and smaller), as can be seen in the bottom panel of Fig. \ref{fig2}, lowering the signal-to-noise ratio of the detection in an unsophisticated way. This parameter space therefore does not require involved investigation. We simulate the galaxies as if they were at the luminosity distances of 50, 100, 150 or 350 Mpc (accordingly lowering the flux and projected size) and inject them into the full volume of our cube.

\subsection{Source finders}
\label{sec:source_finders}

\subsubsection{LESHI}
LESHI\footnote{Leshi (also known as Leshy or Leshen) is a guardian deity of the forest and hunting in Slavic mythology.} (Line Emission Source-Hunting Integrator) is at the core a very straightforward Python script, which is available from github\footnote{\url{https://github.com/misia-mm/LESHI}}. By design, it does not have the sophistication of other source finders, making it simpler to use and install, at the cost of reduced functionality, as it does not attempt any source characterisation and focuses solely on source finding. We note, however, that the exact capabilities of LESHI might change, as it is undergoing active development. As it is relatively fast, it can be used as a first round of source finding to be followed by more advanced tools. LESHI was designed based on the source finding methods done by eye, mimicking its simple tests and checks, as visual source finding has been proven to still be one of the most accurate methods (\citealt{2025A&A...696A.113T}).

\begin{figure}
    \centering
    \includegraphics[width=\linewidth]{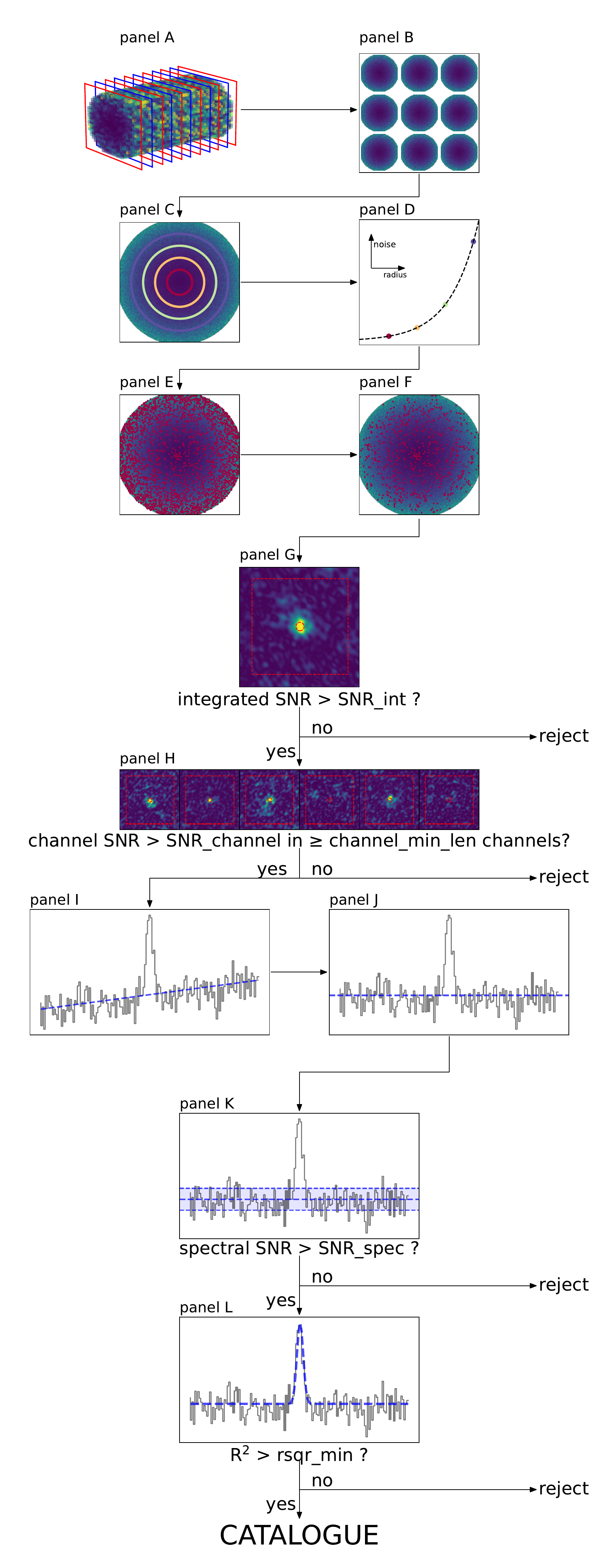}
    \caption{Flowchart visualising each step of the LESHI script. See the main text for more details.}
\label{flowchart}
\end{figure}

The source finder works as follows. As its name suggests, it relies greatly on signal-enhancing integration. First, the cube is divided into thin slabs in frequency of width defined by the {\fontfamily{qcr}\selectfont int\_image\_len} input (panel A on Fig. \ref{flowchart}), which are then integrated, summing all the signals from each channel, creating a number of images (panel B). This is then repeated on the cube divided into thin slabs with frequency moved by half-slab width, to ensure that no weak signal, that happened to be divided into two separate slabs at first, would be missed. To estimate the background noise for each integrated image, the script samples the noise using elliptical apertures (panel C), with the major and minor axis of the ellipse calculated based on the width and height of the datacube, and fits an exponential function (panel D) to the sampled noise, which gives the estimated noise based on the position  relative to the centre of the image. After estimating background noise, all created images are then subjected to the \textsc{find\_peaks()} Astropy Python package (\citealt{2022ApJ...935..167A}) function, which produces a catalogue of all sources detected on the integrated images. The function finds sources above a single specified threshold, which is set based on the noise level at the centre of the image, leading to many false detections towards the edges, where the noise is higher (panel E), due to the primary beam correction. Then, for every source found, the fitted exponential function (from panel D) is used to guess the noise level around the source based on its position. This method is faster than checking the local noise level for every detection candidate (or for a grid), however it is not fault-proof, since it assumes elliptical symmetry, which is not always the case (especially for mosaicked data). We treat it therefore as a first round of higher-tolerance filtering, to mostly weed out the false detections at the edges of the data (panel F). Then, for every detection that has passed, we estimate the local background more accurately by using a box of size specified by the {\fontfamily{qcr}\selectfont bg\_box\_size} input parameter around the source and check if the emission's SNR on the integrated image passes the threshold specified by the {\fontfamily{qcr}\selectfont SNR\_integ} input parameter (panel G). The catalogue of sources that have passed this test is still greatly contaminated by noise peaks, however by investigating each source in the frequency space, those can be identified and excluded. 

The first test in the frequency space checks if at the coordinates of the initially detected source, there is a signal greater than a specified threshold, that persists across channels and is at least of the size equal to the synthesised beam (panel H). If the source is not detected above an SNR threshold specified by the {\fontfamily{qcr}\selectfont SNR\_channel} input in a number or more channels (with the number being specified by the {\fontfamily{qcr}\selectfont channel\_min\_len} input), it is excluded. Next, the spectrum is created from the integrated spectral profile of central pixels within the beam's FWHM at the coordinates of the detected source (panel I) and the script attempts to subtract the spectrum baseline if the {\fontfamily{qcr}\selectfont sloped\_baseline} parameter is set to true (panel J). The second test calculates the SNR of the spectrum (panel K), if the SNR of a given source is below the threshold specified by the {\fontfamily{qcr}\selectfont SNR\_spec} input, it is excluded. Lastly, the code fits a Gaussian function to the spectrum (panel L), using the \textsc{emcee} (\citealt{2013PASP..125..306F}) package, since a real spectral line is expected to have Gaussian-like shape within the beam size. If the detected spectral line does not fit the Gaussian within a  minimum value of coefficient of determination R$^2$ specified by the {\fontfamily{qcr}\selectfont rsqr\_min} input, it is excluded. Each source is then cross-matched and associated with sources that are within a specified number of channels (given by the {\fontfamily{qcr}\selectfont max\_dist\_channel} input) and angular separation (given by the {\fontfamily{qcr}\selectfont max\_dist\_pix} input). Then the final catalogue of the sources that passed all the checks is outputted. 

The LESHI script is parallelised and the number of cores used can be modified using the {\fontfamily{qcr}\selectfont core\_no} input parameter. Its runtime is relatively low (see Sec. \ref{sec:sf_comparison}) and can be further reduced by using more cores. Its memory footprint is also relatively low and only depends on the spatial size of the datacube (not the frequency length), since the script works on a number of integrated images at a time (specified by {\fontfamily{qcr}\selectfont int\_image\_load\_no} input). The memory footprint can be therefore further reduced by lowering the {\fontfamily{qcr}\selectfont int\_image\_load\_no} input parameter, at the cost of increased runtime.

It is worth noting that, since the last Gaussian-fitting test is the most rigorous one, the earlier tests' purpose is to mainly weed out the sample as much as possible, as they are much less computationally expensive than the last test. It is therefore important to find a good combination of input thresholds for the earlier tests to allow through weaker sources, but not too many of them to not make the runtime prohibitively high. A good combination of input parameters, that we have used in this work, is given in Appendix \ref{leshi_sf_params}.

\subsubsection{SoFiA, ProFound and PyBDSF}
\label{sec:sofia}
\textsc{SoFiA} (\citealt{2021MNRAS.506.3962W}, \citealt{2015MNRAS.448.1922S}) is a flexible software application for the detection and parametrization of sources in data cubes.  SoFiA presents the choice of a variety of techniques for data filtering and 3D source-finding and has been developed to be independent of the type of emission line data used. It utilizes the smooth and clip algorithm by iteratively applying spatial and spectral smoothing to the cube on different scales and clipping emission below a certain threshold at each smoothing level. Currently, this is the most commonly-used pipeline for
source finding in \HI\ emission data. \textsc{SoFiA}'s input parameters we use are given in Section \ref{params_SoFiA}.

\textsc{ProFound} (\citealt{2018MNRAS.476.3137R}) is a source finding and image analysis package for two-dimensional data written in the R programming language. It provides tools to generate segmentation maps to discern blended sources and perform photometry. It was mainly developed for optical and infra-red images, but has proven to work well for all types of image data, including radio continuum (\citealt{2019MNRAS.487.3971H}). The input parameters used here for \textsc{ProFound} are given in Section \ref{params_ProFound}.

\textsc{PyBDSF} (Python Blob Detector and Source Finder, \citealt{2015ascl.soft02007M}) is a source-finding Python package developed for radio interferometric two-dimensional data, and was adopted for MIGHTEE continuum data (\citealt{2025MNRAS.536.2187H}). It decomposes images into sets of Gaussians, shapelets or wavelets, and has tools to measure the point-spread function (PSF) variation across an image and calculate spectral indices and polarization properties of sources. \textsc{PyBDSF}'s input parameters used in this work are given in Section \ref{params_PyBDSF}.

The choice of parameters has big impact on the performance of the source finders, as they can be tuned for different uses cases. In this work we use either default or sensible parameters aiming to optimise the completeness and reliability.

\textsc{PyBDSF} and \textsc{ProFound} are two-dimensional source finders, however their capabilities can be extended into three dimensions with a cross-matching step. Both source finders were first run on every image of each channel of our data cube creating a catalogue of detections. Then, to find the sources, every detection was cross-matched with the others using \textsc{STILTs} (\citealt{2005ASPC..347...29T}, \citealt{2006ASPC..351..666T}) and detections within 8 arcseconds radius (which is approximately equal to the beam radius) and one channel away from each other were associated, since we expect real sources to persist in neighbouring channels at the velocity resolution of the cube. Detections that were present in three or more channels were accepted and the rest were rejected as probable noise peaks, since we would not expect any real sources with spectral width of less than three channels, which is $\sim$16.5 km/s for our data.

\section{Simulation results}
\label{sec:results}
\subsection{Completeness}

\begin{figure*}
    \centering
    \includegraphics[width=0.98\linewidth]{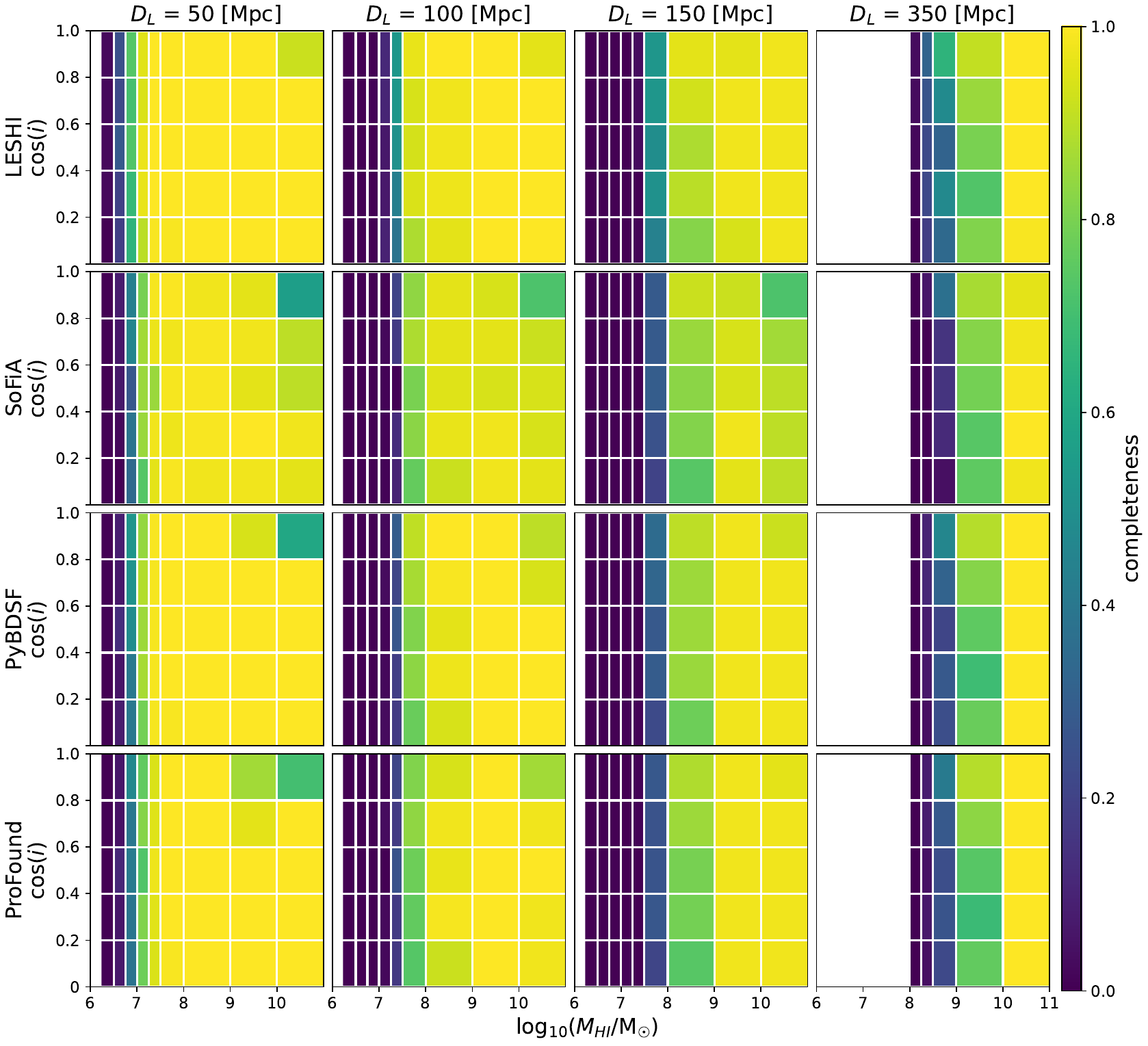}
    \caption{An array of two-dimensional histograms colour-coded by the completeness for each bin of mass (x-axis) and cosine of inclination (y-axis) of injected galaxies, achieved by each of the LESHI, SoFiA, PyBDSF and ProFound source finders (accordingly first, second, third and fourth row of histograms) for each simulated luminosity distance of 50, 100, 150 and 350 Mpc (accordingly first, second, third and fourth column of histograms).  }
\label{completeness_histogram}
\end{figure*}

\begin{figure*}
    \centering
    \includegraphics[width=0.92\linewidth]{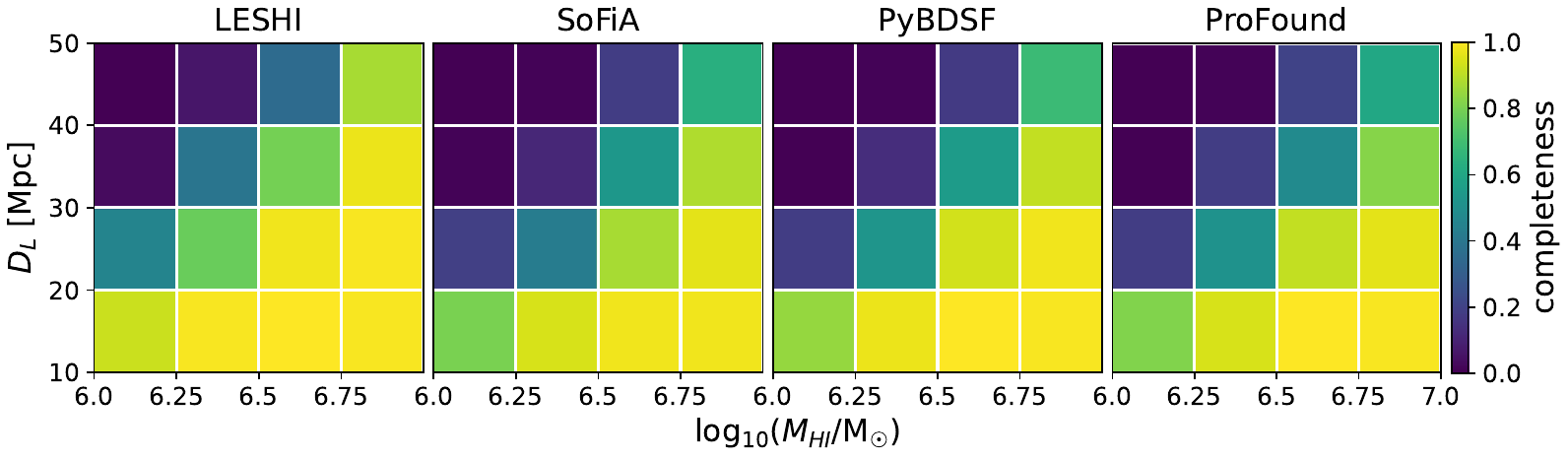}
    \caption{An array of two-dimensional histograms colour-coded by the completeness for each bin of mass (x-axis) and luminosity distance (y-axis) of injected galaxies, achieved by each of the LESHI, SoFiA, PyBDSF and ProFound source finders (accordingly first, second, third and fourth histogram).}
\label{close_by}
\end{figure*}

After completing injection and source finding for each investigated parameter-space bin, we have computed the completeness that each source finder has, by dividing the number of found sources by the number of injected sources, which can be seen in Figures \ref{completeness_histogram} and \ref{close_by}. As anticipated, mass and distance have the greatest influence on the number of sources found, shifting the detection threshold, with the completeness dropping from 1 to 0 within approximately 1 dex of mass. Inclination, although to a much lesser degree, also has an impact, which can be seen in Fig. \ref{completeness_histogram} particularly for the distance of 350 Mpc and the mass bin of $10^{9}-10^{10}$M$_{\odot}$: all source finders perform better for more face-on galaxies, as expected. However, inclination has less impact at smaller distances, where the detection threshold shifts to lower mass galaxies, for which the rotation velocity is much smaller (as a result of Tully-Fisher relation that we adopt) and width of the spectral line is less affected by inclination. Another interesting inclination feature is that for the highest mass bin ($10^{10}-10^{11}$M$_{\odot}$) and face-on galaxies (cos($i$) range of $0.8-1.0$), all source finders are under-performing, with this effect worsening for smaller distances. This is due to the galaxies at these masses and distances having large projected angular sizes, and for face-on inclination all of the emission appears in very few channels, covering a large region (as opposed to each channel showing only a fraction of the area of the galaxy, as can be seen in the middle panel of Fig. \ref{fig2}). If the emission region is of comparable size to the defined regions used for calculating the noise level, it is overestimated and the galaxy is missed. On the other hand, the background region size should not be too large, as this can lead to over-smoothing of local background variations. This effect should be kept in mind when initialising the source finders. It is important to note that PyBDSF and SoFiA have an option for an adaptive box size, that changes its size based on measured local brightness, however this greatly increases computational time.

As the completeness drops to zero for masses below 10$^{7}$ M$_{\odot}$ in the nearest distance bin in Fig. \ref{completeness_histogram}, we have separately explored the parameter space of \HI\ masses 10$^{6} - 10^{7}$ M$_{\odot}$ and distances 10 - 50 Mpc for a constant inclination of 45\textdegree, since inclination is much less important for low-mass galaxies, which can be seen in Fig. \ref{close_by}. The lowest mass galaxies are found only for the distances $<$ 30 Mpc.

\subsection{Completeness function}
\label{sec:completeness_function}
\begin{figure}
    \centering
    \includegraphics[width=\linewidth]{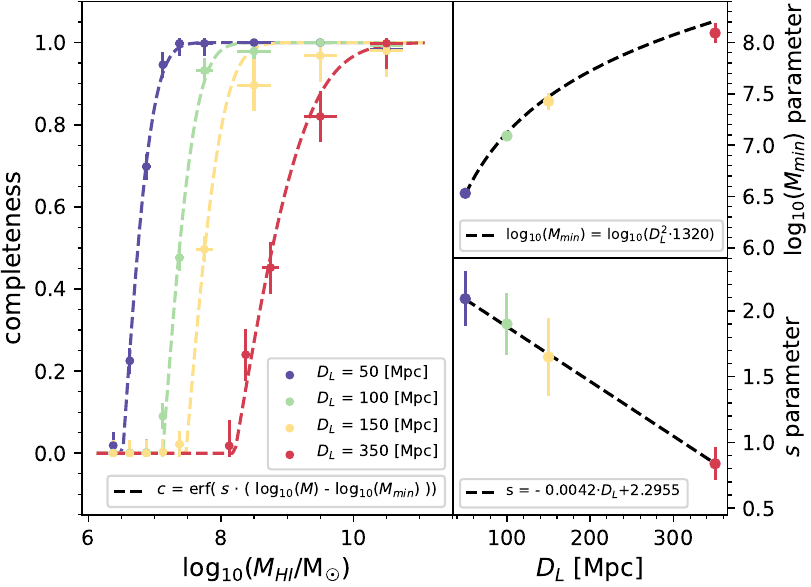}
    \caption{Left panel: inclination averaged completeness achieved by the LESHI source finder \textit{vs} the injected \HI\ mass for different luminosity distances, with the best-fitting error function represented by the dashed lines. Right panels: best-fitting parameters characterising the fitted error function \textit{vs} the luminosity distance.}
\label{completeness_function}
\end{figure}

As expected, completeness depends strongly on the mass and distance of the galaxies. To quantify this relation for the LESHI source finder, we averaged the completeness over all inclinations for given luminosity distance and mass bins, and show this as a function of mass for the four different luminosity distances. We plot the errors accounting for the number of injected sources, assuming Poisson statistics, and errors accounting for the width of the investigated mass bins, calculated by taking the plus minus quarter of the width of the bin (left panel of Fig. \ref{completeness_function}). For each luminosity distance, we fit an error function to the completeness data in the form of:

\begin{equation}
\label{eq:completeness}
  c=\begin{cases}
    \textrm{erf}(s\cdot(\textrm{log}_{10}(\frac{M_{HI}}{\textrm{M}_{\odot}})-\textrm{log}_{10}(\frac{M_{min}}{\textrm{M}_{\odot}}))),&\text{for $M_{HI}>M_{min}$}\\
    0,&\text{otherwise}
  \end{cases}
\end{equation}

\noindent where erf($z$) is the error function defined by erf($z$) = $\frac{2}{\sqrt{\pi}}\int_{0}^{z}e^{-t^{2}}dt$, $c$ is the completeness, $M_{HI}$ is the \HI\ mass, $s$ and $M_{min}$ are free parameters that depend on the luminosity distance: $s$ is responsible for the slope, while $M_{min}$ is the minimal detectable mass ($M_{HI}$ for which the completeness function reaches zero). These functions are overplotted in the left panel of Fig. \ref{completeness_function}.

Next, we plot the best fitting $M_{min}$ and $s$ parameters against the luminosity distance, which can be seen in the right panels of Fig.  \ref{completeness_function}. Since the minimal detectable mass should be proportional to the luminosity distance squared, to find the relation between $M_{min}$ and the distance, we fit a function of the form:
\begin{equation}
\label{eq:slope}
M_{min}(D_{L}) = a_1 \cdot D_{L}^{2}
\end{equation}
where $D_{L}$ is the luminosity distance in Mpc and we find the best fitting value for the parameter $a_1$ to be $1320\pm50$ [$\textrm{M}_{\odot} \textrm{Mpc}^{-2}$] (top right panel of Fig. \ref{completeness_function}). 
Next, we plot the best fitting $s$ parameter for a given luminosity distance against that distance (bottom right panel of Fig. \ref{completeness_function}), and we observe that their relation is very well fitted by a linear function of the form:
\begin{equation}
s(D_{L}) = a_2 \cdot D_{L}+b_2
\end{equation}
and we find the best fitting parameters to be $a_2 = - 0.0042 \pm0.0007$ [Mpc$^{-1}$] and $b_2 = 2.30 \pm 0.18$ (unitless).

It is important to note that a decreasing linear function will eventually reach zero. Therefore, for the fitted $a_2$ and $b_2$ parameters, the $s$ parameter (characterising the slope of the completeness function) would reach zero for the distance of $\sim550$ Mpc, which is not physical, as we expect to detect \HI\ sources at this distance. Therefore, while the function describing how $s$ parameter changes with distance looks linear for the distance range we have investigated, we would expect its true form to asymptotically reach a certain value at high distances (but never zero). To find the function, a similar study might be conducted in the future for higher distances.

\begin{figure}
    \centering
    \includegraphics[width=\linewidth]{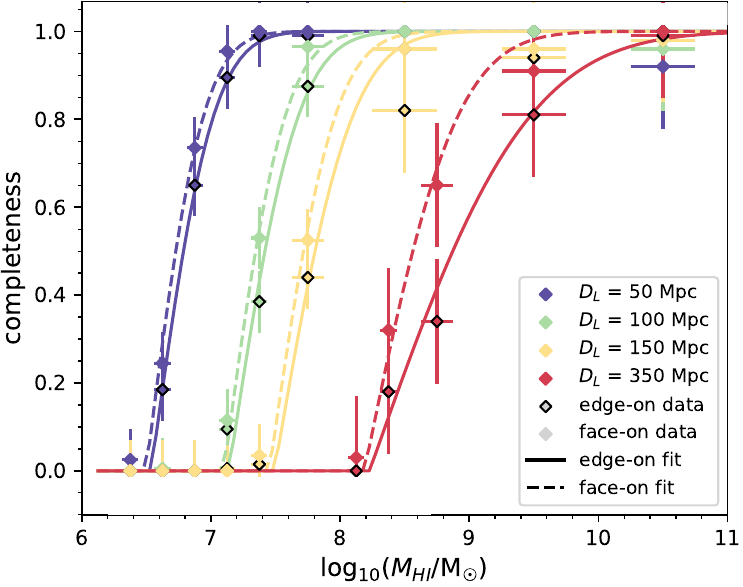}
    \caption{Completeness achieved by the LESHI source finder \textit{vs} the injected \HI\ mass for different luminosity distances for face-on sources (represented by coloured diamonds) and edge-on sources (represented by coloured diamonds with black edge) and the best-fitting error functions represented by the dashed lines (for face-on sources) and solid lines (for edge-on sources).}
\label{completeness_function_ef}
\end{figure}

\begin{table}
\renewcommand{\arraystretch}{1.2} 
\caption{Fitted parameters of the completeness function for the completeness averaged over all inclinations, for face-on sources (cos($i$) range of 0.9 to 1.0) and for edge-on sources (cos($i$) range of 0.0 to 0.1).}
\label{tab:params}
\begin{tabular*}{\linewidth}{@{\extracolsep{\fill}}lccc@{}}
\hline 
 & $a_1$ & $a_2$ & $b_2$ \\
 & {[}$\textrm{M}_{\odot} \textrm{Mpc}^{-2}${]} & {[}Mpc$^{-1}${]} &  \\ \noalign {\smallskip}  \hline\noalign {\smallskip}
average & $1320\pm50$ & $-0.0042\pm0.0007$ & $2.30\pm0.18$ \\
face-on & $1260\pm90$ & $-0.0027\pm0.0014$ & $2.21\pm0.20$ \\
edge-on & $1380\pm110$ & $-0.0039\pm0.0008$ & $2.10\pm0.17$ \\ \noalign {\smallskip} \hline
\end{tabular*}
\end{table}

We have attempted to find a form of the completeness function depending on the inclination, however the sample size was not good enough to find a well constrained relation. Instead, we study the impact of the inclination by considering the two extreme cases: face-on galaxies (cos($i$) range of 0.9 to 1.0) and edge-on galaxies (cos($i$) range of 0.0 to 0.1). We follow the same procedure described above to fit for the parameters describing the completeness function for each case, which can be found in Table \ref{tab:params}. Figure \ref{completeness_function_ef} (analogously to left panel of Fig. \ref{completeness_function}) shows the completeness data and the fitted function for edge-on and face-on sources. We can clearly see that, while the $M_{min}$ is very comparable for the two cases, the slope is lower for the edge-on sources, especially for the distance of 350 Mpc, effectively lowering the completeness and confirming what we have already found from Fig. \ref{completeness_histogram}. Analogous figures and equations for the SoFiA, ProFound and PyBDSF source finders can be found in Appendix \ref{comp_func}.

All the fits were done using the \textsc{emcee} (\citealt{2013PASP..125..306F}) Python package. We used 32 walkers, 5000 steps and burn in phase of 200 steps to discard and a Gaussian likelihood function.

\subsection{Source finder comparison}
\label{sec:sf_comparison}
All source finders could have been optimised differently to satisfy particular use cases. They could be tuned to be more complete at the expense of reliability and vice versa. Therefore, the following results are not representative of the source finders as whole, but rather of the particular setup settings used. In this work we have run the source finders with parameters found through attempting to optimise completeness and reliability for the given data. The input parameters for all source finders and their justification can be found in Appendix \ref{params}.  

Figure \ref{reliability_fragmentation_runtime} showcases a comparison between the source finders: their average completeness (counting only the runs where at least one source finder had at least 5\% completeness), average number of outputted false positives (characterising reliability), average fragmentation (defined as the average number of found sources output by the source finder that belonged to one injected source) and average runtime of the source finders. The number of false positives did not change between the runs, since it should not depend on the injected sources, but on the quality of data and its noise, which was the same for all the runs (since the same real cube has been used).

As can be seen in Fig. \ref{reliability_fragmentation_runtime}, all of the source finders have comparable completeness, with LESHI having the highest value. When it comes to purity, SoFiA performs best, detecting fewest false positives, followed closely by LESHI. SoFiA has the best source fragmentation performance, proving its source characterising capabilities, while LESHI often detects emission belonging to the same galaxy as separate sources, especially for high-mass extended sources, with large projected sizes. Finally, when it comes to average runtime, LESHI was the fastest. It is important to note that all of the investigated source finders can use parallelization. In this work we have run the source finders on a computer cluster with 40 cores. Figure \ref{venn_diagram} shows how many of the injected sources were detected by each source finder and their overlap. Notably, LESHI has the highest number of injected sources detected by no other source finder, while missing only 1.4 per cent of the sources that were detected by at least one source finder. 

\begin{figure}
    \centering
    \includegraphics[width=\linewidth]{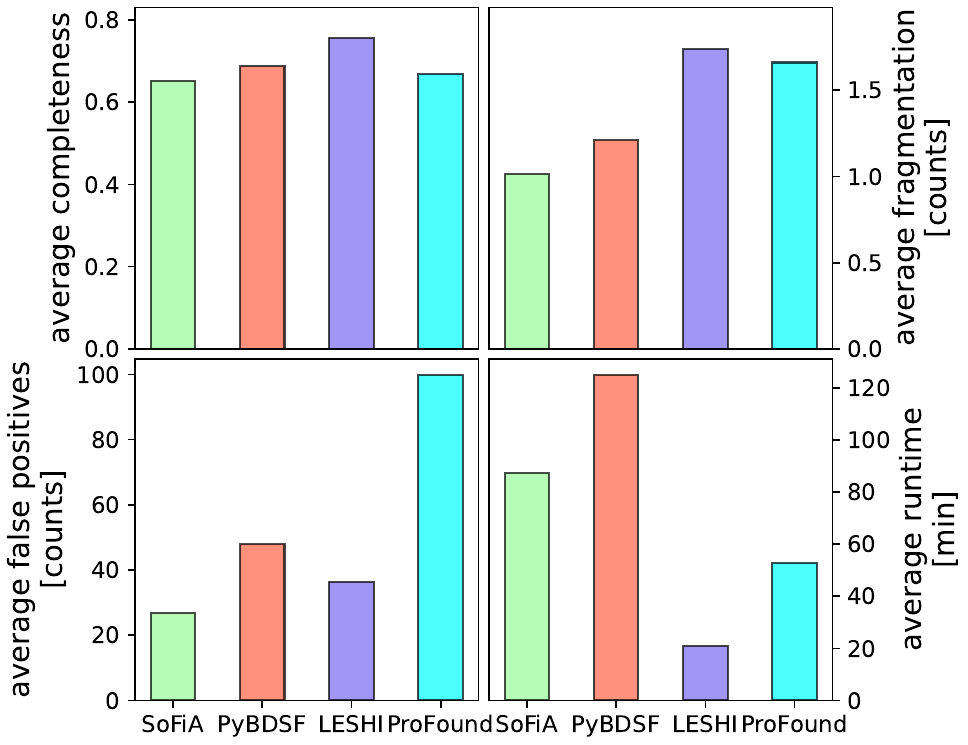}
    \caption{Comparison between the source finders, top left panel shows the average completeness achieved by each source finder, bottom left panel shows the average number of false positives output by each source finder, top right panel shows the average fragmentation of each source finder (defined by the number of sources that were assigned to a single source on average), and lastly, bottom right panel shows the average runtime of each source finder in minutes.}
\label{reliability_fragmentation_runtime}
\end{figure}

\begin{figure}
    \centering
    \includegraphics[width=\linewidth]{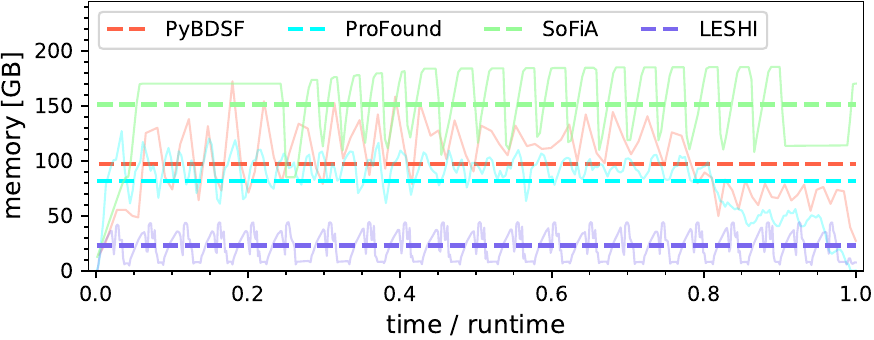}
    \caption{Memory footprints of each source finder run on the 89.3 GB data cube marked in colour, with their average marked by the dashed lines.}
\label{fig:memory}
\end{figure}
\begin{figure}
    \centering
    \includegraphics[width=0.75\linewidth]{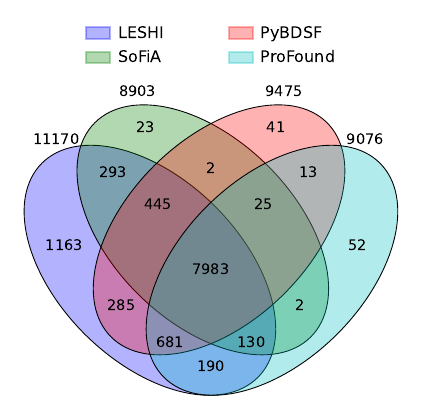}
    \caption{Venn diagram of injected sources that were found by each source finder, representing the overlap between the different source finders (11328 sources were found by at least one source finder out of total of 24000 injected sources). }
\label{venn_diagram}
\end{figure}

Figure \ref{fig:memory} shows the memory usage of each source finder run on the 89.3 GB data cube. It is worth noting that since PyBDSF, ProFound and LESHI work on the channel images separately (with LESHI working on the integrated images), their memory footprints are mostly influenced by the number of parallel threads used, rather than the frequency range of the data cube. Their memory footprint can be therefore adjusted based on the available resources.

While SoFiA is very versatile and can achieve excellent results in the hands of an expert, it has to be optimised differently for each data set. LESHI offers an alternative option, at the expense of source characterisation, and is straightforward to set up even for non-expert users. A collaborator not involved in LESHI's development ran it on different (with respect to pointing centre, spectral resolution, and depth) MeerKAT datasets with recommended settings. This additional testing achieved good results, with new, previously missed, and convincing sources reported by LESHI (private communication).


\section{MIGHTEE-HI DR1 \HI\ catalogue}
\label{sec:catalogue}
In the second part of this work, we perform \HI\ source finding with LESHI and produce a catalogue of new \HI\ sources from the MIGHTEE survey. We measure the \HI\ and stellar properties of the galaxies and compile them into a publicly available catalogue.

\subsection{The MIGHTEE survey}
\label{MIGHTEE}
The MIGHTEE (\citealt{2016mks..confE...6J}) survey is a medium-deep, radio continuum, spectral line and polarisation survey, covering total area of 20 deg$^{2}$ spread over four well-studied extragalactic fields: COSMOS, XMM-LSS, ECDFS and ELAIS-S1. It uses the MeerKAT radio telescope  (\citealt{2009IEEEP..97.1522J}, \citealt{2016mks..confE...1J}), which is located in South Africa and consists of 64 offset-Gregorian dishes, each comprising of a 13.5 m diameter main reflector, covering the baselines up to 8 km. Three receivers operate in the UHF-band, L-band and S-band. The science goals of the MIGHTEE survey focus on three primary aspects: radio continuum (\citealt{2022MNRAS.509.2150H}, \citealt{2025MNRAS.536.2187H}), polarisation (\citealt{2024MNRAS.528.2511T}), and spectral lines (\citealt{2021A&A...646A..35M}).

This work is part of the MIGHTEE-HI working group which focuses on neutral hydrogen sources. The \HI\ sources from the Early Science data (\citealt{2021MNRAS.508.1195P}) were found using unguided visual source finding, covering areas in the COSMOS and XMMLSS fields. In this work we use the LESHI source finder to expand the \HI\ catalogue within the COSMOS field to lower SNR sources and characterise completeness.

The data used in this work is from the first Data Release (DR1) of the MIGHTEE-HI project (see \citealt{2024MNRAS.534...76H} for more details). It covers the COSMOS field and the frequency range 1290--1420 MHz within the L-band, corresponding to 0.0 < z$_{\textsc{HI}}$ < 0.1 for \HI. The data is summarised in Table \ref{tab:my-table}.

\begin{table}
\caption{Brief description of the MIGHTEE-\HI\ DR1 data used in this paper.}
\label{tab:my-table}
\begin{tabular}{ll}
\hline
Area covered            & $\sim$ 4 deg$^{2}$                           \\
Observation time        & 94.2 hours                                  \\
Right ascension range   & $\sim$ 148.89 - 151.17 deg                     \\
Declination range       & $\sim$ 1.20 - 3.23 deg                         \\
Frequency range         & 1290 - 1520 MHz (z = 0 - 0.1 for HI)                            \\
Channel width           & 26.126 kHz (5.5 km s$^{-1}$ at z=0)         \\
Pixel size              & 2 arcsec                                    \\
Median synthesised beam & 15.53 arcsec x 15.53 arcsec  \\                             &(ranging 14.74 - 16.33 arcsec) \\
Sky rms                 & 70 $\mu$Jy/beam at the centre of the mosaic \\
                        & up to 480 $\mu$Jy/beam toward the edge            \\ \hline
\end{tabular}
\end{table}

\subsection{Ancillary data}
The COSMOS field is very well studied and has a plethora of available ancillary data. To confirm our \HI\ detections, we have cross-matched the catalogue with the Sloan Digital Sky Survey sources (SDSS; \citealt{2015ApJS..219...12A}) and The Dark Energy Spectroscopic Instrument catalogue (DESI; \citealt{2025arXiv250314745D}). Photometry for the \textit{g, r, i, z, y} optical bands was measured using the Hyper Suprime-Cam images (HSC; \citealt{2019PASJ...71..114A}, \citealt{2018PASJ...70S...1M}). Photometry for \textit{Y, J, H, K} infrared bands was measured from UKIRT Infrared Deep Sky Survey (UKIDSS; \citealt{2007MNRAS.379.1599L}) Large Area Survey images, which uses the UKIRT Wide Field Camera (WFCAM; \citealt{2007MNRAS.379.1599L}). We note that the COSMOS field has also available UltraVISTA data (\citealt{2012A&A...544A.156M}) for the infrared bands, however our \HI\ data extend beyond the UltraVISTA coverage footprint, therefore we have used the UKIDSS photometry to make consistent measurements. We note that the COSMOS field is also covered by the DEVILS survey (\citealt{2018MNRAS.480..768D}, \citealt{2025MNRAS.544.3005D}), however, while very complete, it only targets the very centre of the field of view of our datacube.

\subsection{Source finding on real data}
\label{sec:source_finding}
Catalogue sources were first identified through untargeted source finding with LESHI with setup parameters given in Appendix \ref{leshi_sf_params}. Every source was then cross-matched with SDSS and DESI catalogues with sources within 8 arcseconds (which is approximately equal to the radius of the beam) and redshift difference less than 0.001 (corresponding to 1.3 MHz or 285 km/s at the median redshift of 0.05), for systems with spectroscopic redshifts, or within redshift errors for systems with photometric redshifts., producing a sample of 530 detections with an optical counterpart. All sources with optical counterparts, integrated SNR greater than five and outside of frequency range with many artifacts ($\gtrsim1306$ MHz) were accepted. Sources with optical counterparts, integrated SNR of 3-5 and/or at frequencies with many artifacts ($\lesssim1306$ MHz) were checked by eye. Sources with the spatial distribution of the emission not consistent with the optical counterpart and with the spectral line being surrounded by strong noise peaks and artifacts resulting from RFI were either excluded or flagged as dubious (see for example Fig. \ref{fig:low_confidence}). All sources were crossmatched with radio continuum data (\cite{2025MNRAS.536.2187H}) and their coordinates checked for a "returning emission" in frequency characteristic of continuum artifacts to rule out their possibility. After removing duplicates (arising from overlapping frequency slabs or over-fragmentation of sources), this resulted in 292 \HI\ sources with plausible optical counterparts and consistent redshifts. The spatial and frequency distribution of all sources can be seen on Fig. \ref{fig:cat_map}.

We have also performed source finding on all the data cubes with the SoFiA, ProFound and PyBDSF source finders, to check if they find any sources that were missed by LESHI. We filtered the candidates following the procedure described above and crossmatched them with the sources found by LESHI. SoFiA found 99 sources, PyBDSF found 155 sources and ProFound found 168 sources with SoFiA finding one additional lower SNR source (see Fig. \ref{fig:only_sofia}). The sources missed by the three source finders (see for example Fig. \ref{fig:only_leshi_det}) were predominantly lower mass, where (as shown by Figures \ref{completeness_histogram} and \ref{close_by}) LESHI's completeness is higher. We note that different settings for all of the above may lead to higher completeness, but introduce lower reliability. Our final catalogue combines the 292 sources detected with LESHI and the one additional source found by SoFiA, totalling in 293 sources.

While the COSMOS field has a wealth of ancillary data facilitating targetted source finding, not every galaxy in the field has optical spectroscopic redshift measured. Therefore, we perform untargeted source finding, which also made it possible to detect any \HI\ emission without an optical counterpart, such as free-floating dark \HI\ clouds. Those detections were noted, however not included in the catalogue.

\subsection{HI mass}
\label{sec:HI_mass}

To measure the total \HI\ flux, we have employed a different approach to the one previously used for MIGHTEE-HI galaxies in \cite{2021MNRAS.508.1195P}, which involved manually smoothing each cubelet and generating an emission mask at a specified sigma level. With increasing numbers of sources, we are aiming to minimise manual intervention. Instead, to extract the emission volume, we start with plotting a spectrum for the central pixels within the beam size at the coordinates of the detection, as this is the minimum spatial size of the emission for unresolved galaxies. We identify the \HI\ line and measure its width by fitting a busy function (described in Section \ref{sec:HI_width}), then we create a moment-0 map by integrating all the channels within the measured width of the spectral line, as determined by the busy function fit. Next, we define the 3$\sigma$ contour on the initial moment-0 map and generate a spectrum from the integrated flux within that contour.  We repeat these steps of generating a spectrum, fitting the busy function, creating a moment-0 map, and generating a new spectrum until the contour changes by less than 5\% of the contour area and all the emission is captured by the found contour within the measured channel width of the global profile.

Using the resulting contour and line profile width, we create a 3-dimensional source mask which we apply to the data cubelet, masking out all pixels that do not belong to the source. We then measure the total \HI\ emission flux by summing the emission of the masked cube. We subtract the local continuum level found through fitting the busy function and we correct the measured flux for the size of the synthesized beam and channel width resulting in flux values  in the units of Jy Hz. To convert from flux density pixel values in the units of Jy beam$^{-1}$ to total measured flux in Jy Hz, we have used the formula:
\begin{align}
\label{eq:conv_units}
S_{tot}=\sum_{pix}^{N_{pix}} S_{\nu,pix}\cdot\delta\nu\cdot\frac{\delta w^{2}}{\frac{\pi \cdot b_{min} \cdot b_{maj}}{4\ln(2)}}
\end{align}

\noindent where $S_{tot}$ is the total measured flux in the units of Jy Hz, $N_{pixel}$ is the number of pixels in the detection volume, $S_{\nu,pix}$ is the flux density for each pixel in Jy beam$^{-1}$, $\delta\nu$ is the width of the channels of the data in Hz, $\delta w$ is the angular size of one pixel (spatial pixel resolution of the data), $b_{maj}$ is beam major axis and $b_{min}$ is beam minor axis, all in consistent units to cancel out. 

To estimate the errors on the \HI\ flux measurement, we create a 7 by 7 grid of spatial coordinates surrounding the source location and central frequency, and offset spatially by the source contour diameter. We apply the source 3-dimensional mask to the data at these offset coordinates and frequency range equal to the frequency range of the source and measure the total flux within the masks, acquiring a set of 49 measured total flux values: one for the source in the centre and 48 for the background around the source. We then calculate the sigma-clipped standard deviation of these 49 values and we take one standard deviation as the error of the measured total flux. We also compute an integrated SNR of the sources by taking the ratio of the measured source total flux and the calculated one standard deviation.

To calculate the total \HI\ mass from the measured flux for each galaxy, we use the formula from \cite{2017PASA...34...52M}, assuming optically thin \HI\ gas:
\begin{equation}
\label{eq:mass}
\left(\frac{M_{HI}}{\text{M$_{\odot}$}}\right) = 49.7\left(\frac{D_{L}}{\text{Mpc}}\right)^{2}\left(\frac{S_{tot}}{\text{Jy Hz}}\right) 
\end{equation}
\noindent where $M_{HI}$ is the total \HI\ mass in units of solar masses, $D_{L}$ is the luminosity distance in Mpc, and $S_{tot}$ is the total observed flux in Jy Hz (=$10^{-26}$W m$^{-2}$). We note that $D_{L}$ depends on the redshift, which in turn greatly depends on assumed peculiar motions of the emission source, especially for local galaxies (\citealt{2008ApJ...676..184T}). Here, we make the simplifying assumption that the spectroscopic redshift represents the distance, ignoring effects from bulk flows. We provide the flux measurements in redshift independent units of Jy Hz, if more accurate calculation of the \HI\ mass are desired. 

We test this mass-measuring method on injected galaxies and compare the measured \HI\ masses to the true ones. As can be seen on the left panel of Figure \ref{inj_comparison}, our method accurately measures the \HI\ mass within the estimated uncertainties.

\subsection{HI profile widths}
\label{sec:HI_width}
To characterise the \HI\ emission spectra and measure their velocity widths in an automated manner, we fit a busy function (Equation 4 in \citealt{2014MNRAS.438.1176W}) to the profile of the spectrum, as done in \cite{2021MNRAS.508.1195P}. The function is given by:
\begin{equation}
\begin{split}
B(x)&=\frac{a}{4}\times(\erf(b_{1}\times(w+x-x_{e}))+1) \\
&\times(\erf(b_{2}\times(w-x+x_{e}))+1)\times(c\times\left| x-x_{p} \right|^{2}+1)+C
\end{split}
\end{equation}
\begin{figure}
    \centering
    \includegraphics[width=\linewidth]{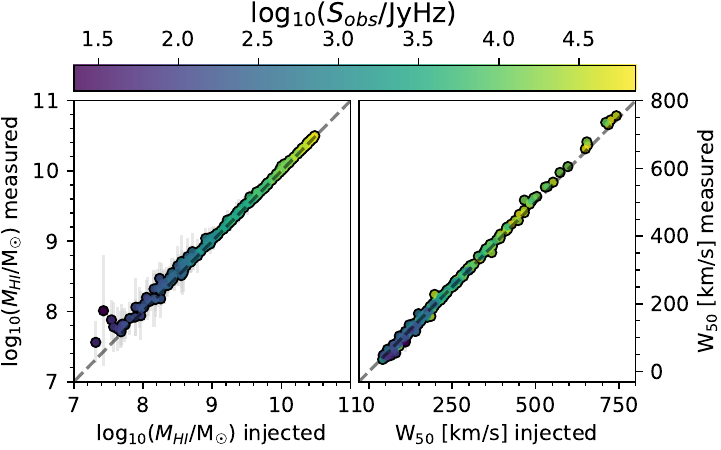}
    \caption{Comparison between \HI\ mass (left panel) and velocity width (right panel) of measured emission (y-axes) and injected emission (x-axes), colour-coded by the observed \HI\ flux.}
\label{inj_comparison}
\end{figure}
where erf(z) is the error function defined by $\frac{2}{\sqrt{\pi}}\int_{0}^{z}e^{-t^{2}}dt$ and $a$, $b_1$, $b_2$, $x_e$, $x_p$, $w$, $c$, $C$ are the free parameters changing the exact form of the function\footnote{The busy function can be explored using the Desmos online graphing calculator: \url{https://www.desmos.com/calculator/bbc9ed8be4}}. The function is able to fit well even the most complicated double-horned profiles (see bottom left panel of Fig. \ref{example_galaxy}), but at the cost of many free parameters. To fit for the eight free parameters we use the \textsc{emcee} Python package, using 32 walkers, 5000 steps and burn in phase of 500 steps that were discarded and a Gaussian likelihood function. We then measure the velocity width W$_{50}$ at 50\% of the average peak flux density of the fitted function, found by identifying the two peaks of the spectral profile and calculating their average value. Since none of the fitted parameters are solely responsible for the width of the spectral profile, we cannot use the output posterior plots to estimate the error on the width. Instead, we measure the widths for each fitted function tried out by the sampler and calculate the error of W$_{50}$ from the 16th and 84th percentiles of the histogram of the measured widths. Some of the output errors are very small, so we combine this error in quadrature with the velocity width of the channels (calculated for the given redshift), equal to $\sim$ 5.5 km s$^{-1}$ for $z=0$, accounting for the velocity resolution of our data.

We test this method on injected galaxies and compare the measured velocity widths to the true ones. As can be seen on the right panel of Figure \ref{inj_comparison}, our method estimates the width within the measurement uncertainties.

\begin{figure*}
    \centering
    \includegraphics[width=1\linewidth]{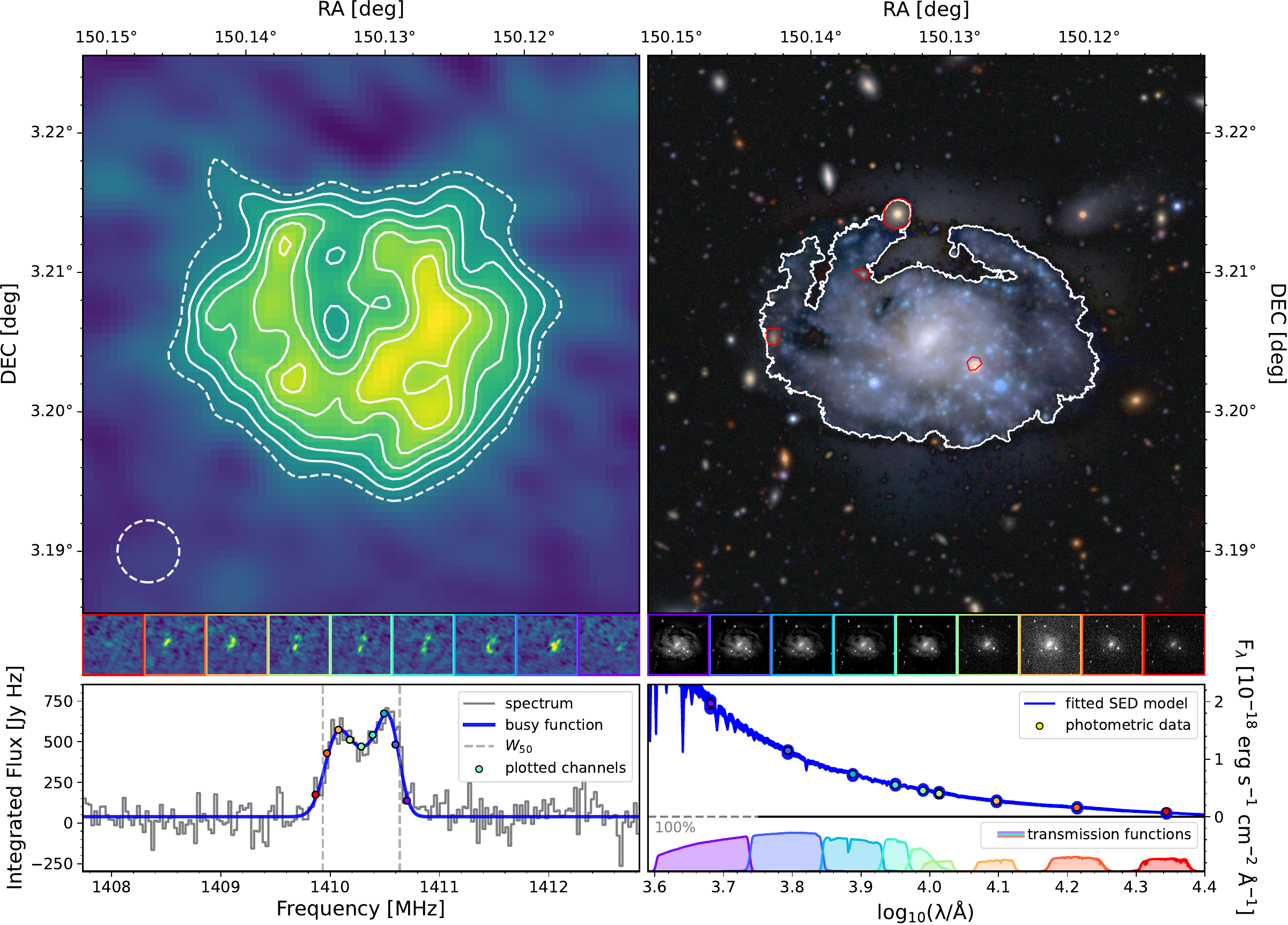}
    \caption{Top left: \HI\ moment-0 map with contours over-plotted on top for an example galaxy from our sample, marking column densities of 3.0, 4.3, 5.7, 7.0, 8.4, 9.7 x $10^{20}$ cm$^{-2}$ with the size of the synthesised beam marked by a dashed circle. The row of 9 channel maps at the bottom show the \HI\ emission for the individual channels, which are marked by points on the spectrum below with their colour matching the border of the channel map. Bottom left: Observed \HI\ emission spectrum (marked in gray) with fitted busy function (marked in blue) with the measured velocity width W$_{50}$ marked by vertical dashed lines. The spectrum is for the integrated flux within the first contour on moment-0 map (marked by dashed line) tracing 3$\sigma$ emission. Top right: Hyper Supreme Cam $gri$ composite optical image for the example galaxy from our sample with g-band flux contours from DRUID (\citealt{2025RASTI...4....6S}) in white, and contours surrounding rejected components marked in red. The row of 9 images at the bottom show the emission for each of the \textit{g, r, i, z, y}, \textit{Y, J, H, K} photometric bands. Bottom right: multi-wavelength flux measurements for photometric bands (coloured points) with the best-fit SED model from BAGPIPES (\citealt{2018MNRAS.480.4379C}) marked in blue, and the transmission functions for each of the \textit{g, r, i, z, y}, \textit{Y, J, H, K} photometric bands plotted below in colour.}
\label{example_galaxy}
\end{figure*}

\subsection{Optical photometry and SED fitting}
\label{sec:photometry}
The galaxies from our sample are at relatively low redshift, meaning they are very well resolved in optical images, complicating the measurement of their photometry. To extract the source fluxes in an automated way, we have used the  Detector of astRonomical soUrces in optIcal and raDio images (DRUID) source detection software (\citealt{2025RASTI...4....6S}), which utilises persistent homology to detect sources and their components. We have utilised DRUID here as it performs very well at associating complex morphology sources while keeping track of substructure, which often characterises clumpy, star-forming, well-resolved galaxies, like the ones from our sample. We perform the initial photometry on the HSC \textit{g}-band images, as our galaxies tend to be blue and actively star-forming. To extract the galaxy and reject any background/foreground sources, each detected component was compared against the local flux level and colour of the main galaxy by measuring their brightness and colour from the emission in the \textit{g} and \textit{i} band images within the contour of the component. If its brightness differed by more than specified value (a factor of 2 for sources redder than the main galaxy and a factor of 10 for sources bluer than the main galaxy), it was rejected. An example contour can be seen on the upper right panel of Fig. \ref{example_galaxy} with rejected components marked in red. Every galaxy has been then checked by eye, and the contours modified if needed. The contours were then used to create a mask, which was applied to each of the \textit{g, r, i, z, y}, \textit{Y, J, H, K} photometric band images and the emission within the mask was summed and background subtraction applied. 
To convert the raw pixel values from the band images into magnitudes, we have used the following formula:

\begin{align}
m_{obs}=-2.5\cdot \textrm{log}_{10}(S_{tot})+m_{zp}+m_{AB}
\end{align}

\noindent where $m_{obs}$ is the observed magnitude in the AB magnitude system, $S_{tot}$ is the total pixel value for the HSC images and  the total pixel value divided by the exposure time for the UKIDSS images, $m_{zp}$ is the magnitude zeropoint equal to 27 for the HSC images, and for the UKIDSS images the zeropoint values were taken from the header for each band image, $m_{AB}$ is the magnitude offset to convert magnitudes between the AB and Vega system for the UKIDSS, taken from \cite{2006MNRAS.367..454H}.

To calculate the errors of the measured flux we have used the following formula:

\begin{align}
\sigma^{2}=\frac{1}{g}\sum_{n}^{N_{A}}(S_{n}-\overline{B}) + N_{A}\cdot\sigma^{2}_{B/pix}
\end{align}

\noindent where $\sigma$ is the calculated error, $g$ is the gain of the detector in electrons per pixel data unit, $S_{n}$ is signal in pixel n in image data units, $\overline{B}$ is estimated median background per pixel, $N_{A}$ is number of pixels in source aperture and $\sigma_{B/pix}$ is pixel standard deviation of the sky background. Following \cite{2021MNRAS.506.4933A} and \cite{2023MNRAS.524.4586V} we adopt a minimum flux uncertainty of 5\% to mitigate against issues around mismatch between the synthetic galaxy templates and the observations of real galaxies, whilst also accounting for colour-dependent zero-points for the individual filters. To convert the errors to magnitude system we have used the standard formula: 

\begin{align}
\Delta m_{obs}=\frac{2.5}{\textrm{ln}10}\cdot \frac{\Delta S_{tot}}{S_{tot}}
\end{align}

\noindent where $\Delta m_{obs}$ is the error in magnitudes and $\Delta S_{tot}$ is the measured flux error. For the errors taken to be equal to 5\% of the measured flux, their value in magnitude system is equal to the constant value of 0.054.

To obtain estimates of the stellar masses and star formation rates, we employ the Bayesian Analysis of Galaxies for Physical Inference and Parameter Estimation code (BAGPIPES; \citealt{2018MNRAS.480.4379C}) to perform spectral energy distribution (SED) fitting based on our measured multi-wavelength fluxes. The code uses a Chabrier Initial Mass Function (\citealt{2003PASP..115..763C}), coupled with Stellar Population Synthesis models based on \cite{2003MNRAS.344.1000B} with several dust extinction and star formation history models. In this work we use the delayed star formation history model along with the dust attenuation model using the Calzetti law (\citealt{2000ApJ...533..682C}), by allowing the extinction to vary between $A_{V} = 0-2$. During the fitting, we keep the redshift fixed to that measured from the \HI\ emission. The bottom right panel of Fig. \ref{example_galaxy} shows an SED fit for an example galaxy from our sample. The results of the fitting can be found in Table \ref{tab:main_table}, with the uncertainties based on the posterior distributions output by BAGPIPES calculated from the 16th and 84th percentiles.

The inclination was estimated by measuring the axis ratio of isophotes fitted with the \textsc{photutils} Python package (\citealt{larry_bradley_2025_14889440}) to the emission in the \textit{g}-band and derived using the standard relation:

\begin{equation}
\textrm{cos}(i)^{2}=\frac{\left(\frac{b}{a}\right)^2-q_{0}^{2}}{1-q_{0}^{2}}
\end{equation}
\noindent where $\frac{b}{a}$ is the axis ratio of the fitted outermost isophote and $q_{0}$ is the intrinsic axis ratio of the disc (with typical values between 0 and 0.4; \citealt{1990ApJ...349....1F}), which we assumed to be equal to 0.2. Each isophote ellipse was checked by eye and corrected where needed.

\begin{table*}
\caption{Table containing first 42 columns from the catalogue for a sample of eight galaxies; five omitted columns contain flags with values of either 0 or 1. See the main text for more details. The full table is available as supplementary material for this paper.}
\label{tab:main_table}


\begin{tabular}{llllllllllll}
\hline
ID catalogue & RA & Dec & Freq. & z$_{\text{HI}}$ & z$_{\text{spec}}$ & D$_{\text{L}}$ & S$_{\text{HI}}$ & $\delta$S$_{\text{HI}}$ & log$_{10}$(M$_{\text{HI}}$) & $\delta$log$_{10}$(M$_{\text{HI}}$) \\
 & {[}deg{]} & {[}deg{]} & {[}MHz{]} &  &  & {[}Mpc{]} & {[}Jy Hz{]} & {[}Jy Hz{]} & {[}M$_{\odot}${]} & {[}M$_{\odot}${]}  \\ \hline
MGTH\_J100256.4+023440 & 150.735 & 2.578 & 1360.267 & 0.0442 & 0.0442 & 195.6 & 1471.09 & 109.80 & 9.45 & 0.03  \\
MGTH\_J095642.7+022509 & 149.178 & 2.419 & 1306.161 & 0.0875 & 0.0874 & 399.1 & 202.78 & 35.36 & 9.21 & 0.08 \\
MGTH\_J095951.4+014224 & 149.964 & 1.707 & 1385.660 & 0.0251 & 0.0249 & 109.3 & 1074.15 & 68.53 & 8.81 & 0.03  \\
MGTH\_J100006.8+022245 & 150.028 & 2.379 & 1403.974 & 0.0117 & 0.0117 & 50.6 & 193.29 & 16.14 & 7.39 & 0.04  \\
MGTH\_J095741.1+020149 & 149.422 & 2.031 & 1376.726 & 0.0317 & 0.0317 & 138.7 & 132.05 & 22.94 & 8.10 & 0.08 \\
MGTH\_J100156.7+030737 & 150.486 & 3.127 & 1321.157 & 0.0751 & 0.0751 & 339.7 & 596.37 & 125.30 & 9.54 & 0.09 \\
MGTH\_J095914.8+021131 & 149.812 & 2.192 & 1386.287 & 0.0246 & 0.0245 & 107.4 & 703.72 & 70.68 & 8.61 & 0.04  \\
MGTH\_J100357.1+022505 & 150.988 & 2.418 & 1383.623 & 0.0266 & 0.0266 & 116.7 & 12782.92 & 672.47 & 9.93 & 0.02  \\ \hline
\end{tabular}

\begin{tabular}{llllllllllllll}
\hline
SNR$_{\textrm{3D}}$ & W$_{50}$ & $\delta$W$_{50}$ & incl. & axis ratio & log$_{10}$(M$_{\text{stel}}$) & $\delta$log$_{10}$(M$_{\text{stel}}$) & SFR & $\delta$SFR & m$_g$ & $\delta$m$_g$ & m$_r$ & $\delta$m$_r$\\
 & {[}km/s{]} & {[}km/s{]} & {[}deg{]} &  & {[}M$_{\odot}${]} & {[}M$_{\odot}${]} & {[}M$_{\odot}$/year{]} & {[}M$_{\odot}$/year{]} & {[}mag{]} & {[}mag{]} & {[}mag{]} & {[}mag{]}\\\hline
13.4 & 167 & 7 & 44 & 0.73 & 9.11 & 0.07 & 0.24 & 0.12 & 18.37 & 0.05 & 17.96 & 0.05\\
5.9 & 191 & 20 & 41 & 0.77 & 9.59 & 0.10 & 1.65 & 0.79 & 18.18 & 0.05 & 17.91 & 0.05\\
16.0 & 101 & 7 & 61 & 0.51 & 7.54 & 0.12 & 0.07 & 0.03 & 19.22 & 0.05 & 19.14 & 0.05\\
12.0 & 40 & 6 & 49 & 0.67 & 6.75 & 0.11 & 0.00 & 0.00 & 20.32 & 0.05 & 20.07 & 0.05\\
5.4 & 100 & 16 & 57 & 0.57 & 8.48 & 0.08 & 0.05 & 0.03 & 19.25 & 0.05 & 18.85 & 0.05\\
4.4 & 280 & 27 & 59 & 0.54 & 10.10 & 0.10 & 3.20 & 1.78 & 18.03 & 0.05 & 17.41 & 0.05\\
10.3 & 103 & 8 & 71 & 0.37 & 7.54 & 0.11 & 0.02 & 0.01 & 20.32 & 0.05 & 20.02 & 0.05\\
19.1 & 392 & 12 & 75 & 0.32 & 10.33 & 0.10 & 4.44 & 2.64 & 15.07 & 0.05 & 14.47 & 0.05\\ \hline
\end{tabular}

\begin{tabular}{llllllllllllllll}
\hline
m$_i$ & $\delta$m$_i$ & m$_z$ & $\delta$m$_z$ & m$_y$ & $\delta$m$_y$ & m$_Y$ & $\delta$m$_Y$ & m$_J$ & $\delta$m$_J$ & m$_H$ & $\delta$m$_H$ & m$_K$ & $\delta$m$_K$ \\
{[}mag{]} & {[}mag{]} & {[}mag{]} & {[}mag{]} & {[}mag{]} & {[}mag{]} & {[}mag{]} & {[}mag{]} & {[}mag{]} & {[}mag{]} & {[}mag{]} & {[}mag{]} & {[}mag{]} & {[}mag{]} \\ \hline
17.75 & 0.05 & 17.64 & 0.05 & 17.50 & 0.05 & 17.98 & 0.09 & 18.35 & 0.18 & 17.40 & 0.10 & 17.84 & 0.19 \\
17.66 & 0.05 & 17.61 & 0.05 & 17.46 & 0.05 & 17.73 & 0.05 & 17.54 & 0.06 & 17.49 & 0.07 & 17.58 & 0.09 \\
19.14 & 0.05 & 19.09 & 0.05 & 19.00 & 0.05 & 18.81 & 0.17 & 18.67 & 0.17 & 19.45 & 0.35 & 26.93 & 99 \\
20.03 & 0.05 & 19.95 & 0.05 & 19.90 & 0.05 & 21.73 & 1.62 & 19.95 & 0.41 & 21.36 & 1.95 & 20.38 & 0.95 \\
18.64 & 0.05 & 18.52 & 0.05 & 18.40 & 0.05 & 18.77 & 0.16 & 19.48 & 0.38 & 18.59 & 0.16 & 19.14 & 0.39 \\
17.02 & 0.05 & 16.82 & 0.05 & 16.57 & 0.05 & 16.74 & 0.05 & 16.44 & 0.05 & 16.35 & 0.05 & 16.23 & 0.05 \\
19.90 & 0.05 & 19.80 & 0.05 & 19.71 & 0.05 & 20.04 & 0.36 & 19.56 & 0.31 & 20.49 & 0.93 & 19.41 & 0.41 \\
14.17 & 0.05 & 13.93 & 0.05 & 13.70 & 0.05 & 13.75 & 0.05 & 13.64 & 0.05 & 13.40 & 0.05 & 13.54 & 0.05 \\ \hline
\end{tabular}

\end{table*}

\subsection{Multi-wavelength \HI\ source catalogue}

The \HI, optical and NIR photometry, as well as derived parameters for the full catalogue can be found online as explained in Section \ref{data_avail}. Example measurements for a sample of 10 galaxies can be found in Table \ref{tab:main_table}. The table's keywords and description are:
\begin{enumerate}
\itemindent=-0pt
\setlength\itemsep{0.5em}

\item Column 1 - {\fontfamily{qcr}\selectfont ID\_catalogue} - a unique identifier computed based on sky coordinates of each galaxy.
\item Column 2 - {\fontfamily{qcr}\selectfont RA\_deg} - right ascension (J2000) of the optical counterpart of the \HI\ detection, in degrees.
  
  \item Column 3 - {\fontfamily{qcr}\selectfont Dec\_deg} - declination (J2000) of the optical counterpart of the \HI\ detection, in degrees.
  \item Column 4 - {\fontfamily{qcr}\selectfont freq\_MHz} - frequency centred on the \HI\ detection, in MHz.
  \item Column 5 - {\fontfamily{qcr}\selectfont z\_HI} - redshift calculated using the formula:
  \begin{equation}
  z = \frac{\nu_{rest} - \nu_{obs}}{\nu_{obs}}
  \end{equation}
  where $\nu_{obs}$ is the measured frequency of the \HI\ detection and $\nu_{rest}$ is equal to 1420.40575 MHz. 
  \item Column 6 - {\fontfamily{qcr}\selectfont z\_spec} - spectroscopic redshift of galaxies crossmatched from the DESI catalogue, values of -99 mark the galaxies not found in the DESI catalogue.
  \item Column 7 - {\fontfamily{qcr}\selectfont D\_L\_Mpc} - luminosity distance calculated assuming $\Lambda$CDM cosmology with $H_{0} = 70 \text{ km s}^{-1}\text{Mpc}^{-1}$, $\Omega_{\text{M}}=0.3$ and $\Omega_{\Lambda}=0.7$ and ignoring any possible cosmic flows, in Mpc.
  \item Column 8 (9) - {\fontfamily{qcr}\selectfont S\_HI\_Jy\_Hz} ({\fontfamily{qcr}\selectfont S\_HI\_Jy\_Hz\_err}) - total observed flux (and its error) of the detected \HI\ emission, in Jy Hz. For full description of the measurement method see Section \ref{sec:HI_mass}.
  \item Column 10 (11) - {\fontfamily{qcr}\selectfont log\_M\_HI} ({\fontfamily{qcr}\selectfont log\_M\_HI\_err}) - base 10 logarithm of the measured total \HI\ mass (and its error) in the units of solar masses, calculated from the total \HI\ flux using Equation \ref{eq:mass}.
  \item Column 12 - {\fontfamily{qcr}\selectfont SNR\_3D} - integrated SNR of the \HI\ detection, calculated as described in Section \ref{sec:HI_mass}.
  \item Column 13 (14) - {\fontfamily{qcr}\selectfont W\_50\_km\_s} ({\fontfamily{qcr}\selectfont W\_50\_km\_s\_err}) - \HI\ detection spectral profile velocity width (and its error) measured at 50\% of the flux peak, in km s$^{-1}$. Note that it is not inclination corrected. For full description of the measurement method see Section \ref{sec:HI_width}.
  \item Column 15 - {\fontfamily{qcr}\selectfont incl\_deg} -  inclination of the optical disk, in degrees, with edge-on galaxy having an inclination of 90$^{\circ}$ and face-on 0$^{\circ}$, measured as described in Section \ref{sec:photometry}.
  \item Column 16 - {\fontfamily{qcr}\selectfont axis\_ratio} -  ratio between the minor and major axis of the optical disk, measured as described in Section \ref{sec:photometry}.
  \item Column 17 (18) - {\fontfamily{qcr}\selectfont log\_M\_stel} ({\fontfamily{qcr}\selectfont log\_M\_stel\_err}) - base 10 logarithm of the total stellar mass (and its error) in the units of solar masses, derived through SED fitting, as described in Section \ref{sec:photometry}.
  \item Column 19 (20) - {\fontfamily{qcr}\selectfont sfr\_M\_sol\_year} ({\fontfamily{qcr}\selectfont sfr\_M\_sol\_year\_err}) - star formation rate (and its error) in the units of solar masses per year, derived through SED fitting, as described in Section \ref{sec:photometry}.
  \item Columns 21 - 38 - {\fontfamily{qcr}\selectfont m\_g} - {\fontfamily{qcr}\selectfont m\_K}  ({\fontfamily{qcr}\selectfont m\_g\_err} - {\fontfamily{qcr}\selectfont m\_K\_err}) - observed magnitudes (and their errors) of the flux measured for the \textit{g, r, i, z, y}, \textit{Y, J, H, K} optical and infrared bands in the AB magnitude system, as described in Section \ref{sec:photometry}.
  \item Column 39 - {\fontfamily{qcr}\selectfont low\_confidence\_flag} - flag marking low-confidence detections. Sources with flag 0 have high signal-to-noise and there is little doubt that the detection is genuine. Detections with flag 1, while having an optical counterpart, were judged to be unreliable due to their spatial distribution and spectral profile. An example is given in Fig. \ref{fig:low_confidence}. This flag does not disqualify the detection, but reflects lower confidence in it and its measurements.
  \item Column 40 - {\fontfamily{qcr}\selectfont blended\_flag} - flag marking blended sources (flag 1) for which the galaxies have small separation and their \HI\ contents can not be separated and resolved. Sources with flag 0 are isolated sources. An example is given in Fig. \ref{fig:blended}.
  \item Column 41 - {\fontfamily{qcr}\selectfont confused\_flag} - flag marking potentially confused sources (flag 1) for which the galaxies have small separation (similar to the blended sources), however the morphology of the \HI\ distribution does not strongly suggest a blended source.  An example is given in Fig. \ref{fig:confused}.
  \item Column 42 - {\fontfamily{qcr}\selectfont bad\_ellipse\_flag} - flag marking irregularly shaped sources  (flag 1) in the optical for which a confident axis ratio (and therefore inclination) could not be determined.
  \item Column 43 - {\fontfamily{qcr}\selectfont contaminated\_source\_flag} - flag marking galaxies for which the optical photometry is contaminated by foreground/background sources.
\end{enumerate}
To retrieve the "golden sample", the sum of all flags should add up to zero.

\begin{figure}
    \centering
    \includegraphics[width=\linewidth]{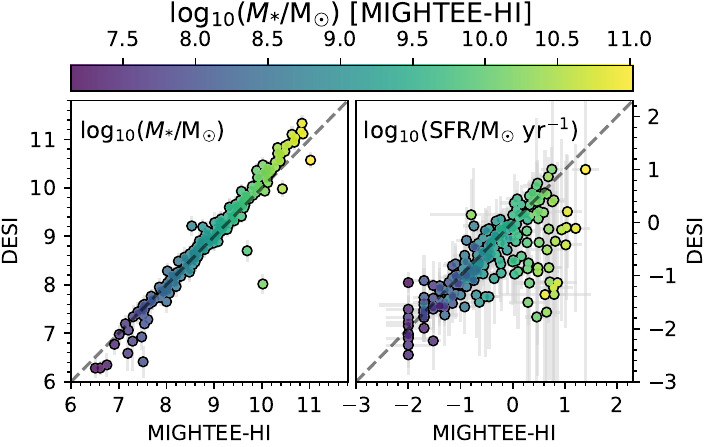}
    \caption{Comparison between stellar mass (left panel) and star-formation rate (right panel), determined through SED fitting in this work (x-axes) and DESI (y-axes) for crossmatched galaxies present in both catalogues.}
\label{desi_comparison}
\end{figure}

In Fig. \ref{desi_comparison} we compare the stellar population properties inferred through SED fitting in this work and in DESI (\citealt{2024ApJ...961..173Z}, \citealt{2025arXiv250314745D}). As can be seen in the two panels, stellar masses and star-formation rates follow the one-to-one relation very closely with the exception of the most massive systems for which DESI masses are slightly higher and star-formation rates lower. This might be attributed to the different measurement methods and photometric bands used. For example, stellar population properties in \cite{2024ApJ...961..173Z} were inferred without the use of near-infrared data and utilising spectroscopy. A similar discrepancy was observed in Fig. 6 in \cite{2024ApJ...961..173Z}, where the DESI stellar masses were on average larger across all masses than the ones inferred in the Cosmic Evolution Survey (COSMOS; \citealt{2022ApJS..258...11W}).

\begin{figure}
    \centering
    \includegraphics[width=\linewidth]{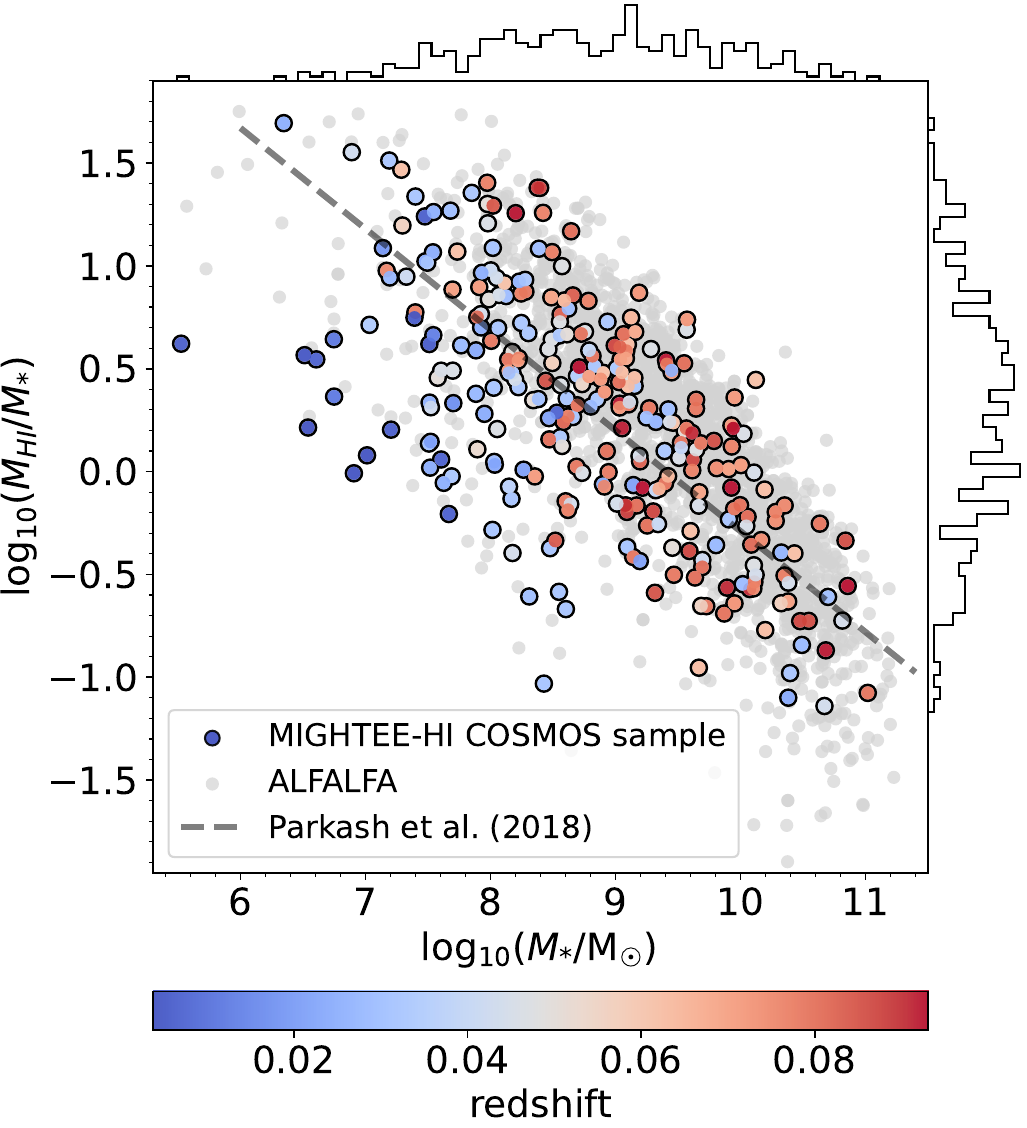}
    \caption{The gas fraction (ratio of \HI\ mass to stellar mass) as a function of the stellar mass, colour-coded by redshift for galaxies from our catalogue and sample distribution histograms plotted on top and on the right of the figure. Galaxies from the ALFALFA survey are plotted in gray. Dashed line marks the relation adapted from Eq. 8 in \citet{2018ApJ...864...40P}.}
\label{galaxies}
\end{figure}

Figure \ref{galaxies} shows the gas fraction (ratio of \HI\ mass to stellar mass), as a function of the stellar mass for the galaxies from our catalogue, colour-coded by their redshift. Our sample agrees well with the relation between \HI\ and stellar masses from Eq. 8 in \cite{2018ApJ...864...40P}, which was derived for \HI\ selected sample of galaxies and suggests that more massive galaxies have lower \HI\ gas fraction. In the figure, we can see that higher redshift galaxies tend to be above the relation, since at those redshifts for a given stellar mass, we will detect the most \HI-rich galaxies.

We compare our derived \HI\ masses to the ones from public catalogue from \cite{2023MNRAS.522.5308P} for galaxies present in both works (see Fig. \ref{fig:es}). The measurements agree well with each other and we note that any differences most likely arise due to the MIGHTEE data used in this work having higher spectral resolution and sensitivity than the Early Science data.

\section{Discussion}

\begin{figure}
    \centering
    \includegraphics[width=0.96\linewidth]{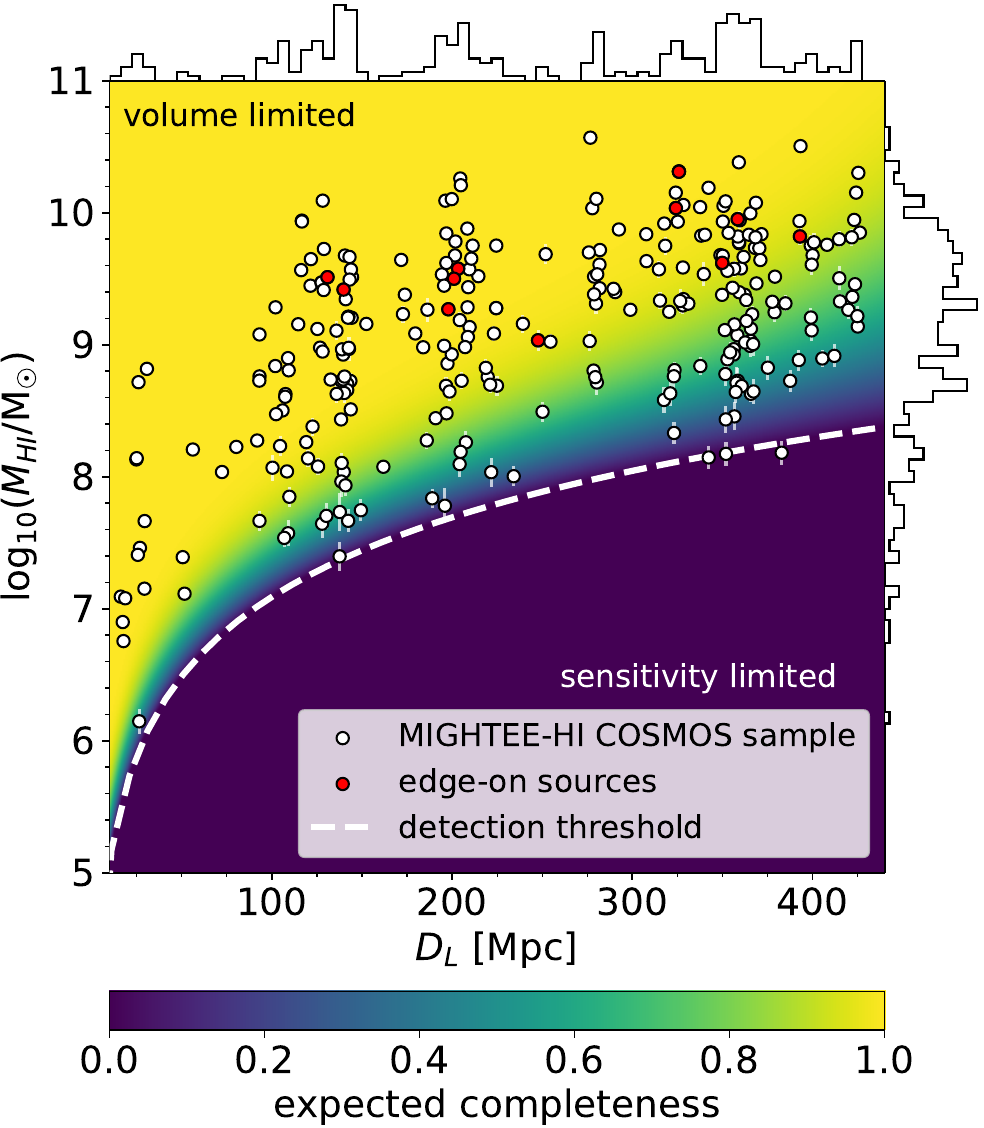}
    \caption{\HI\ mass of the catalogue galaxies plotted against their luminosity distance represented by white points, with most edge-on sources plotted in red and sample distribution histograms plotted on top and on the right of the figure. The theoretical completeness map calculated using Eq. \ref{eq:completeness} is plotted as the background in colour with the detection threshold (where the expected completeness drops to zero given by Eq. \ref{eq:slope}) marked by the white dashed line.}
\label{sample_with_completeness}
\end{figure}

Combining the results from the two main parts of this work, we use Eq. \ref{eq:completeness} for the completeness $c(M_{\text{\HI}},D_{L})$ to plot a completeness map by colour-coding every point in the $M_{\text{\HI}}$-$D_{L}$ parameter space. We then overplot the real detections from our galaxy sample and see how the results from the two parts of this work compare. As can be seen on Fig. \ref{sample_with_completeness}, at all distances and masses, our detections lie above the expected detection threshold, where the derived expected completeness drops to zero. Moreover, most edge-on detections (cos$(i)<0.1$) marked in red, all lie much above the detection threshold, consistent with our findings that high inclination sources are more difficult to detect.

Computing the completeness through source finding on injected sources is critical in deriving the \HI\ mass function by correcting the sample densities based on their estimated completeness. For this purpose, LESHI is ideal as its relatively low runtime and memory footprint allow for many iterations of source finding on simulated sources to reach high number statistics and thus accurate completeness estimates.

However, one of the caveats of our artificial source injection approach is that the galaxies modelled using \textsc{$^{\text{3D}}$Barolo} are idealised; they have symmetric spectra and do not posses any irregularities characteristic of real sources. However, we do not expect it to influence the results by much, as the source finders rely on the emission being above the noise level, rather than it being symmetric or not.

An important thing to note is that source finding done on artificial sources is expected to be more complete and deeper compared to source finding on real data. For injected sources we know exactly where they are and even very low SNR sources will be accepted. For real systems, we need their emission to have large enough SNR to accept them as a genuine. However, this issue is somewhat diminished in our work by the availability of optical data, since even very low SNR sources might be accepted if they have an optical counterpart with consistent spectroscopic redshift.

\section{Summary and conclusions}
\label{sec:conclusions}

In this work, we have introduced a catalogue of 293 \HI\ sources found in the MIGHTEE survey data cubes, through untargeted source finding. The sources lie in the COSMOS field in the redshift range of $0.004<z<0.093$. This area has a wealth of ancillary data available, making it possible to crossmatch each of our \HI\ detections to optical counterparts with measured spectral redshifts, resulting in high expected reliability of our sample. In addition to \HI\ masses and velocity widths of the \HI\ emission lines, the catalogue includes optical through near-infrared photometry and stellar population properties, inferred through SED fitting. As the contents of \HI\ catalogue acquired through untargeted source finding greatly depend on the source finding methods used, this study also provides a well-characterised expected completeness of the detected sample of galaxies based on their properties, inferred through a comparative study of different source finding algorithms.

In the first part of this work, we have compared the SoFiA, ProFound, PyBDSF and LESHI source finders, by injecting a sample of simulated systems into the MIGHTEE data cube. We note that the study's aim was not to compare the capabilities of optimised and maximally fine-tuned source finders, but rather their capabilities our collaboration would be able to realistically achieve, by aiming to find the best balance between the reliability and completeness within reasonable amount of time, for the purposes of catalogue creation. We have found that for the setup settings used, SoFiA had the lowest number of false positives, proving its reliability and did not over-fragment the sources, while LESHI proved to be the most complete (at a cost of slightly worsened reliability) and fastest. PyBDSF and ProFound perform very well even though they were designed for two-dimensional data, with their completeness and reliability being comparable to the completeness of LESHI and SoFiA. LESHI has very low memory footprint equal to around 20 per cent of the size of the datacube on average (with jumps of up to 50 per cent), which can be lowered further depending on the input number of integrated images to be analysed at once, at the expense of a longer runtime. Similarly, since PyBDSF and ProFound work on separate channel images in parallel, their memory footprint can also be adjusted. They can be therefore run on very large data sets without the need of dividing them into subcubes. With this in mind, a possible automated source finding strategy could include a first round performed with one of the fast, low memory footprint, and very complete source finders, followed by targeted source finding with SoFiA to properly characterise the sources. It is important to note that the way each source finder performs depends greatly on the setup settings used. Each could have been made more complete at the expense of reliability and vice versa.

We have also investigated how the completeness of the source finders depends on the inclination, mass and distance of simulated galaxies. We have observed that the inclination has an impact on the completeness, especially for higher distances, where high mass galaxies are at the detection threshold, making it more challenging to find edge-on sources. This is likely due to the fact that higher-mass galaxies rotate faster, and the spectral line broadening for high inclinations (as explained in Section \ref{sec:Inclinations}), is more severe than for lower masses. We then averaged the completeness over the investigated inclinations and quantified the relation between completeness and mass and distance with a functional form (Fig. \ref{completeness_function} and Eq. \ref{eq:completeness}). We have shown that the completeness changes rapidly with mass, changing from 0\% to 100\% in the span of 0.5 dex for lower distances and 1 dex for higher distances. The slower rate of change of the completeness function for higher distances can be attributed to the inclination effect on completeness for higher mass galaxies. The mass where the completeness drops to zero is proportional to the logarithm of the distance squared, consistent with the standard relation between the observed flux and luminosity distance. The constant of proportionality for this relation ($a_{2}$ in Eq. \ref{eq:slope}) should depend on the sensitivity of the data and Eq. \ref{eq:completeness} could be potentially applied to other surveys, after changing that constant accordingly. We have also learnt to be cautious of face-on, high-mass and nearby galaxies, which may be missed by the source finders due to improper estimation of the background level. 

As the era of the SKAO comes closer, the lessons learned can be used in devising source finding strategies for the anticipated amounts of new data. Moreover, the relationships determined between completeness and galaxy properties for the MIGHTEE data can be now used in statistical studies, such as determining the \HI\ mass function.

\section*{Data Availability}
\label{data_avail}
The catalogue is available online as the supplementary material for this paper.

The MIGHTEE-\HI\ spectral cubes are available from \url{https://doi.org/10.48479/jkc0-g916 } (\citealt{2024MNRAS.534...76H}). The optical and near-infrared data images used in this work are all in the public domain (please see the references cited in the main text). Other data underlying the article are available on request to the first author.

\section*{Acknowledgements}
We thank the reviewer for prompt, detailed and insightful comments. We thank Prof. Marc Verheijen for sharing his experience and advice, which greatly contributed to the creation of this paper. This work is based on observations made by the MeerKAT telescope, which is operated by the South African Radio Astronomy Observatory, which is a facility of the National Research Foundation, an agency of the Department of Science and Innovation. We acknowledge use of the InterUniversity Institute for Data Intensive Astronomy (IDIA) data intensive research cloud for data processing. IDIA is a South African university partnership involving the University of Cape Town, the University of Pretoria and the University of the Western Cape. The study described in the paper made use of \textsc{Astropy} (\citealt{2022ApJ...935..167A}), Cube Analysis and Rendering Tool for Astronomy (CARTA; \citealt{2021zndo...4905459C}, \textsc{TOPCAT} \citep{2005ASPC..347...29T}, \textsc{NumPy} \citep{2011CSE....13b..22V}, \textsc{SciPy} \citep{2020zndo...4100507V}, \textsc{matplotlib} \citep{2007CSE.....9...90H} and NASA's Astrophysics Data System.

TGH acknowledges support from UNIQ+ scholarship. MB gratefully acknowledges the financial support from the Flemish Fund for Scientific Research (FWO-Vlaanderen) and the South African National Research Foundation (NRF) under Bilateral Scientific Cooperation program (grant G0G0420N). He also acknowledge the support of networking activities by NRF and the Belgian Science Policy Office (BELSPO) under grant BL/02/SA12 (GALSIMAS). SLJ acknowledges the support of a UKRI Frontiers Research Grant [EP/X026639/1], which was selected by the European Research Council, and the STFC consolidated grants [ST/S000488/1] and [ST/W000903/1]. MG is supported through UK STFC Grant ST/Y001117/1. MG acknowledges support from the Inter-University Institute for Data Intensive Astronomy (IDIA). IDIA is a partnership of the University of Cape Town, the University of Pretoria and the University of the Western Cape. CLH acknowledges support from the Oxford Hintze Centre for Astrophysical Surveys which is funded through generous support from the Hintze Family Charitable Foundation.



\bibliographystyle{mnras}
\bibliography{references} 




\appendix

\section{Source finders input parameters}
\label{params}
In this section we state the input parameters used for each source finder along with the justification for the chosen values of parameters that have an impact on the source finding. We have attempted to find a set of input parameters giving good results for the MIGHTEE data and the best balance between the completeness and reliability.

\subsection{SoFiA}
\label{params_SoFiA}
We have run the SoFiA source finder with the following input parameters:

\begin{verbatim}
# Global settings
pipeline.verbose           =  false
pipeline.pedantic          =  false
pipeline.threads           =  8

# Input
input.data                 = data_cube
input.gain                 =
input.noise                =
input.weights              =
input.mask                 =
input.invert               =  false

# Flagging
flag.region                =
flag.catalog               =
flag.radius                =  5
flag.auto                  =  false
flag.threshold             =  5.0
flag.log                   =  false

# Continuum subtraction
contsub.enable             =  true
contsub.order              =  0
contsub.threshold          =  2.0
contsub.shift              =  4
contsub.padding            =  3

# Noise scaling
scaleNoise.enable          =  true
scaleNoise.mode            =  local
scaleNoise.statistic       =  mad
scaleNoise.fluxRange       =  negative
scaleNoise.windowXY        =  51
scaleNoise.windowZ         =  9999
scaleNoise.gridXY          =  0
scaleNoise.gridZ           =  0
scaleNoise.interpolate     =  true
scaleNoise.scfind          =  false

# Ripple filter
rippleFilter.enable        =  false
rippleFilter.statistic     =  median
rippleFilter.windowXY      =  21
rippleFilter.windowZ       =  21
rippleFilter.gridXY        =  0
rippleFilter.gridZ         =  0
rippleFilter.interpolate   =  false

# S+C finder
scfind.enable              =  true
scfind.kernelsXY           =  0, 8, 16
scfind.kernelsZ            =  0, 3, 5, 7, 15, 31
scfind.threshold           =  3.8
scfind.replacement         =  2.0
scfind.fluxRange           =  negative
scfind.statistic           =  mad

# Threshold finder
threshold.enable           =  false
threshold.threshold        =  1e-30
threshold.mode             =  absolute
threshold.statistic        =  mad
threshold.fluxRange        =  negative

# Linker
linker.enable              =  true
linker.radiusXY            =  2
linker.radiusZ             =  2
linker.minSizeXY           =  8
linker.minSizeZ            =  6
linker.maxSizeXY           =  0
linker.maxSizeZ            =  0
linker.minPixels           =  0
linker.maxPixels           =  0
linker.minFill             =  0.0
linker.maxFill             =  0.0
linker.positivity          =  false
linker.keepNegative        =  false

# Reliability
reliability.enable         =  true
reliability.parameters     =  peak, sum, mean
reliability.threshold      =  0.8
reliability.scaleKernel    =  0.2
reliability.minSNR         =  3.0
reliability.minPixels      =  0
reliability.autoKernel     =  true
reliability.iterations     =  30
reliability.tolerance      =  0.05
reliability.catalog        =
reliability.plot           =  true
reliability.debug          =  false

# Mask dilation
dilation.enable            =  false
dilation.iterationsXY      =  10
dilation.iterationsZ       =  5
dilation.threshold         =  0.001

# Parameterisation
parameter.enable           =  true
parameter.wcs              =  true
parameter.physical         =  true
parameter.prefix           =  SoFiA
parameter.offset           =  true

# Output
output.directory           =  output_dir
output.filename            =  output_file
output.writeCatASCII       =  true
output.writeCatXML         =  true
output.writeCatSQL         =  false
output.writeNoise          =  true
output.writeFiltered       =  true
output.writeMask           =  true
output.writeMask2d         =  true
output.writeRawMask        =  false
output.writeMoments        =  true
output.writeCubelets       =  true
output.marginCubelets      =  10
output.thresholdMom12      =  0.0
output.overwrite           =  true
\end{verbatim}

The input parameters of the SoFiA source finder with an impact on the completeness and reliability are: 
\begin{enumerate}
    \item {\fontfamily{qcr}\selectfont scfind.threshold} - sets the flux threshold used by the smooth+clip source finder in the units of background rms relative to the noise level. Values in the range of 3-5 are recommended in SoFiA's manual. We have set this parameter to 3.8, as we wanted to be sensitive towards weaker sources and which leads to a good compromise between the completeness and reliability.
    \item {\fontfamily{qcr}\selectfont scfind.kernelsXY} - list of spatial Gaussian kernel sizes used to smooth the data in the spatial domain. We have set it to "0, 8, 16" since the synthesised beam's diameter of our data is equal to 8 pixels.
    \item {\fontfamily{qcr}\selectfont reliability.enable} - if set to true then the reliability calculation and filtering will be performed, which determines the reliability of each detection and discards any that are below the specified reliability threshold. We have set it to true following the recommendation from SoFiA's manual, as the value of the {\fontfamily{qcr}\selectfont scfind.threshold} is in the lower range, which leads to good achieved reliability. We have run the source finder with {\fontfamily{qcr}\selectfont reliability.enable} set to false (without changing any other parameters), which led to $\sim7000$ false positives.
    \item {\fontfamily{qcr}\selectfont reliability.threshold} - sets the reliability threshold sources below which are discarded. The default value is 0.9; we have set it to 0.8 to be more tolerant to weaker sources.
\end{enumerate}

\subsection{ProFound}
\label{params_ProFound}
We have run the ProFound source finder with the following input parameters:

\begin{verbatim}
out_pro = profoundProFound(
channel_image, 
plot=FALSE, 
skycut=3.0, 
pixcut=16, 
tolerance= 1, 
ext=1,
box = c(100,100),
rotstats=TRUE, 
boundstats=TRUE, 
nearstats=TRUE,
groupstats=TRUE, 
verbose=FALSE)
\end{verbatim} 
The input parameters of the ProFound source finder with an impact on the completeness and reliability are: 
\begin{enumerate}
    \item {\fontfamily{qcr}\selectfont skycut} - sets the threshold for the object in number of sky rms above the mean. We have tried values in the range of 2-4 and found that 3 leads to a good compromise between the completeness and reliability,
    \item {\fontfamily{qcr}\selectfont pixcut} - sets the minimum number of pixels of an object. We have set it to 16, as this is approximately the area of the synthesised beam of our data and we would expect the emission size to be no smaller than the beamsize.
    \item {\fontfamily{qcr}\selectfont tolerance} - sets the minimum height of the object in number of sky rms above the mean between its highest point and the point where it contacts another object and is responsible for combining different components. Recommended range is 1-5, we have set to 1 to avoid associating very close separate sources.
    \item {\fontfamily{qcr}\selectfont box} - size of the box in pixels used to estimate the background noise. We have set it c(100,100) to not over-smooth any spatial noise variations. Size of 100 pixels is also much larger than most of the sources expected to be found.
\end{enumerate}

\subsection{PyBDSF}
\label{params_PyBDSF}
We have run the PyBDSF source finder with the following input parameters:

\begin{verbatim}
out_pyb = bdsf.process_image(
channel_image, 
thresh_isl = 2.5, 
thresh_pix = 3.0, 
adaptive_rms_box=False,
advanced_opts=True,
group_by_isl=True,
group_method='intensity',
group_tol=1.0,
ini_gausfit='nobeam', 
minpix_isl=16,
peak_fit=True,
rms_value=None,
split_isl=False,
atrous_do=False,
flagging_opts=True,
flag_minsize_bm=1,
flag_minsnr=0.7,
flag_smallsrc=True,
beam=(hdr_im['BMAJ'],hdr_im['BMIN'],hdr_im['BPA']),
frequency=hdr_im['CRVAL3'],
interactive=False,
mean_map='default',
multichan_opts=False,
output_opts=True,
quiet=True,
polarisation_do=False,
psf_vary_do=False,
rms_box=(45,15),
rms_map=None,
shapelet_do=False,
spectralindex_do=False,
thresh='hard'
)
\end{verbatim}

The input parameters of the PyBDSF source finder with an impact on the completeness and reliability are: 
\begin{enumerate}
    \item {\fontfamily{qcr}\selectfont thresh\_isl} - threshold for the island boundary in number of sky rms above the mean. The default value is 3, however we have set it to 2.5, to not reject the weaker sources, which also leads to larger number of false positives, most of which is discarded at the crossmatching with other channels step. We have tried values in the range of 2-4 and found that 2.5 leads to a good compromise between the completeness and reliability.
    \item {\fontfamily{qcr}\selectfont thresh\_pix} - threshold for the island peak in number of sky rms above the mean. The default value is 5, however similarly to {\fontfamily{qcr}\selectfont thresh\_isl} we have set it to the lower value of 3. We have tried values in the range of 2.5-5 and found that 3 leads to a good compromise between the completeness and reliability.
    \item {\fontfamily{qcr}\selectfont minpix\_isl} - minimum number of pixels with emission per island. We have set it to 16, as this is approximately the area of the synthesised beam of our data and we would expect the emission size to be no smaller than the beamsize.
\end{enumerate}

\subsection{LESHI}
\label{leshi_sf_params}
We have run the LESHI source finder with the following input parameters:
\begin{verbatim}
LESHI.source_finder(
data_file,
path_to_results='./',
SNR_integ=3.5, 
SNR_channel=2.5, 
channel_min_len=3,
SNR_spec=3, 
rsqr_min=0.35,
int_image_len=10, 
int_image_load_no=10,
channel_start=0, 
channel_end=None, 
beam=None, 
bg_box_size=100, 
max_dist_pix=10, 
max_dist_channel=10, 
test_hist=True,
sloped_continuum=False,
core_no=40)
\end{verbatim}

The input parameters of the LESHI source finder with an impact on the completeness and reliability are: 
\begin{enumerate}
    \item {\fontfamily{qcr}\selectfont int\_image\_len} - length of the frequency slab to be integrated into a moment-0 map on which the first phase of source finding is performed. We have set it to 10 channels, as this is approximately equal to the width of the weakest expected signals in the MIGHTEE data ($\sim$ 50 km/s).
    \item {\fontfamily{qcr}\selectfont SNR\_integ} - threshold SNR of the source on the integrated image to be accepted. We have set it to 3.5, as this is high enough to filter out most of the noise peaks, but low enough to allow for weak sources.
    \item {\fontfamily{qcr}\selectfont SNR\_channel} - threshold SNR of the source on the single channel image to be accepted, we have set it to 2.5 to allow for any weaker sources.
    \item {\fontfamily{qcr}\selectfont channel\_min\_len} - minimum number of consecutive channels the sources has to persist in (with SNR specified by {\fontfamily{qcr}\selectfont SNR\_channel}) to be accepted, we have set it to 3, as we would not expect any real source to have the width narrower than 3 channels ($\sim$ 15 km/s).
    \item {\fontfamily{qcr}\selectfont SNR\_spec} - threshold SNR of the source in the spectral dimension, we have set it to 3 to filter out most of the noise peaks, while allowing weaker sources.
    \item {\fontfamily{qcr}\selectfont rsqr\_min} - minimum values of the R$^2$ parameter characterising how well a Gaussian function fits to the spectrum of the detection. We have set it to 0.35, as it filters out most of the remaining noise peaks and leaves out the genuine "bumps" in the spectrum, leading to a good compromise between the completeness and reliability. 
\end{enumerate}

\section{Completeness function}
\label{comp_func}
\begin{figure*}
    \centering
    \includegraphics[width=0.50\linewidth]{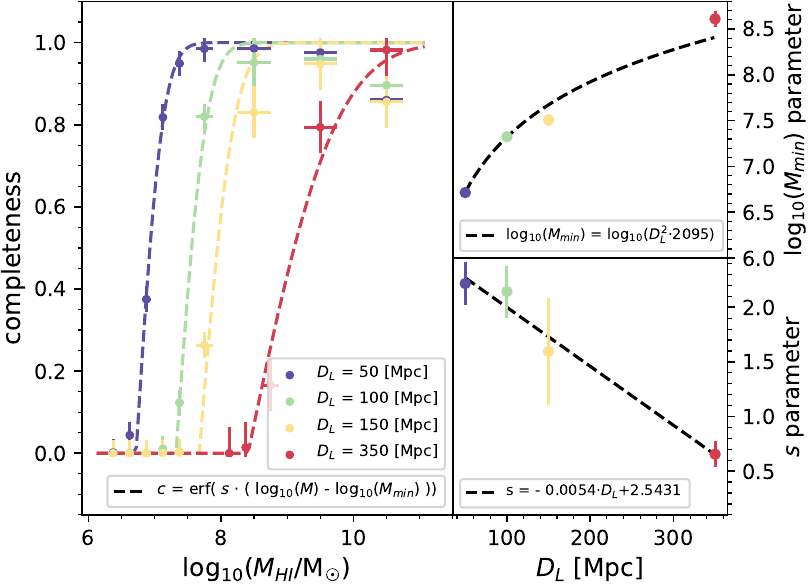}\hfill
    \includegraphics[width=0.46\linewidth]{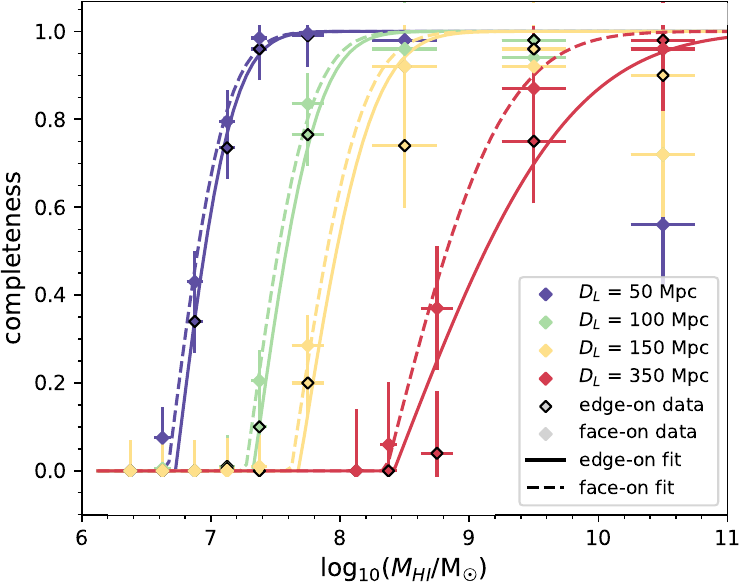}
    \caption{Figures analogous to Fig. \ref{completeness_function} and Fig. \ref{completeness_function_ef} for the SoFiA source finder.}
\label{fig:comp_sofia}
\end{figure*}

\begin{figure*}
    \centering
    \includegraphics[width=0.50\linewidth]{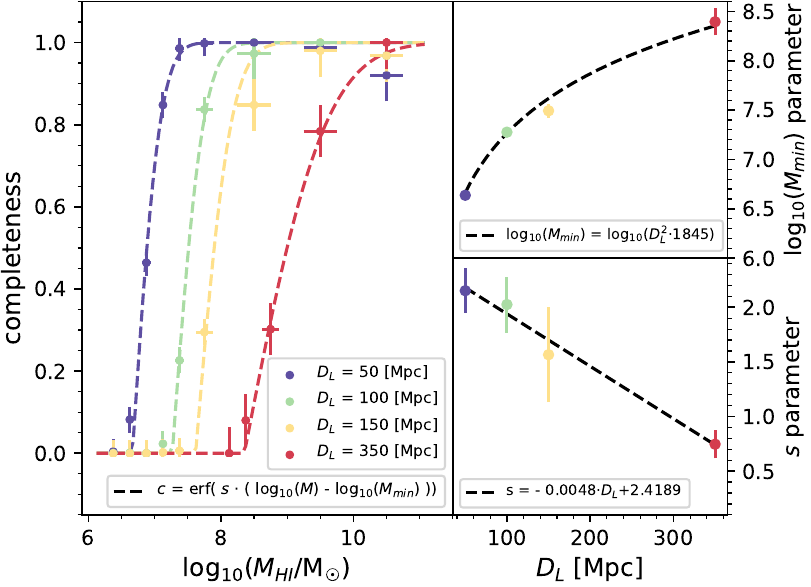}\hfill
    \includegraphics[width=0.46\linewidth]{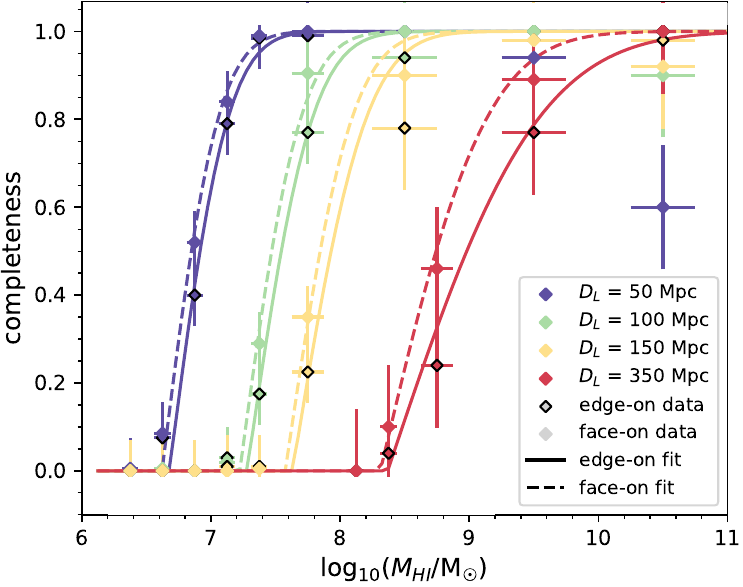}
    \caption{Figures analogous to Fig. \ref{completeness_function} and Fig. \ref{completeness_function_ef} for the PyBDSF source finder.}
\label{fig:comp_pybdsf}
\end{figure*}

\begin{figure*}
    \centering
    \includegraphics[width=0.5\linewidth]{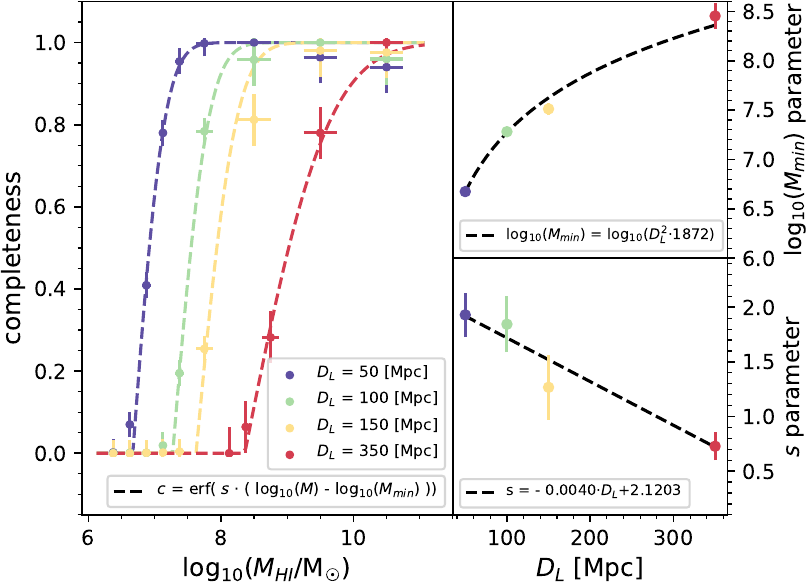}\hfill
    \includegraphics[width=0.46\linewidth]{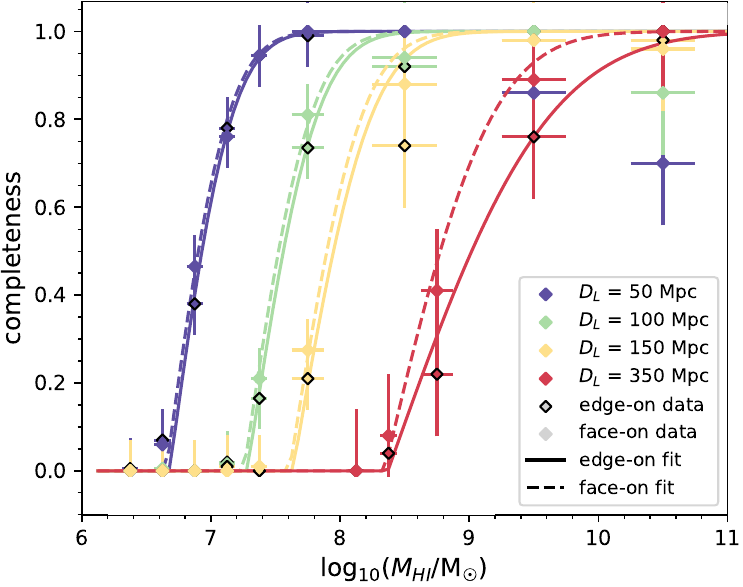}
    \caption{Figures analogous to Fig. \ref{completeness_function} and Fig. \ref{completeness_function_ef} for the ProFound source finder.}
\label{fig:comp_profound}
\end{figure*}

Table \ref{tab:comp_function} contains the parameters describing the fitted completeness function (Eq. \ref{eq:completeness}) for the SoFiA, PyBDSF and ProFound source finders, derived analogously to Section \ref{sec:completeness_function}. Figures \ref{fig:comp_sofia}, \ref{fig:comp_pybdsf} and \ref{fig:comp_profound} contain plots showing the fitted completeness function analogously to Fig. \ref{completeness_function} and Fig. \ref{completeness_function_ef} for the SoFiA, PyBDSF and ProFound source finders. Qualitatively, the fitted completeness functions for all of the sourcefinders are very similar. Fitted $M_{min}$ parameter follows well the $M_{min}\sim D_L^2$ relation and 
the fitted slope parameter $s$ is lower for the edge-on sources, especially for higher distances, making the completeness for highly inclined sources lower, as discussed in the main text.

\begin{table}
\caption{Fitted parameters of the completeness function for the different source finders for the completeness averaged over all inclinations, for face-on sources (cos($i$) range of 0.9 to 1.0) and for edge-on sources (cos($i$) range of 0.0 to 0.1).}
\label{tab:comp_function}
\begin{tabular}{llll}
\hline
Source finder & $a_1$ & $a_2$ & $b_2$ \\
 & {[}$\textrm{M}_{\odot} \textrm{Mpc}^{-2}${]} & {[}Mpc$^{-1}${]} &  \\ \hline
SoFiA &  &  &  \\
average & $2095\pm57$ & $-0.0054\pm0.0007$ & $2.54\pm0.19$ \\
face-on & $1854\pm147$ & $-0.0035\pm0.0011$ & $2.24\pm0.21$ \\
edge-on & $2126\pm167$ & $-0.0048\pm0.0008$ & $2.35\pm0.21$ \\ \hline
PyBDSF &  &  &  \\
average & $1845\pm74$ & $-0.0048\pm0.0007$ & $2.41\pm0.19$ \\
face-on & $1672\pm130$ & $-0.0033\pm0.0012$ & $2.26\pm0.21$ \\
edge-on & $1905\pm165$ & $-0.0041\pm0.0009$ & $2.21\pm0.21$ \\ \hline
ProFound &  &  &  \\
average & $1872\pm77$ & $-0.0040\pm0.0007$ & $2.12\pm0.19$ \\
face-on & $1800\pm146$ & $-0.0028\pm0.0011$ & $2.05\pm0.20$ \\
edge-on & $1913\pm178$ & $-0.0039\pm0.0008$ & $2.10\pm0.17$ \\ \hline
\end{tabular}
\end{table}

\section{Catalogue sources map}
\begin{figure*}
    \centering
    \includegraphics[width=\linewidth]{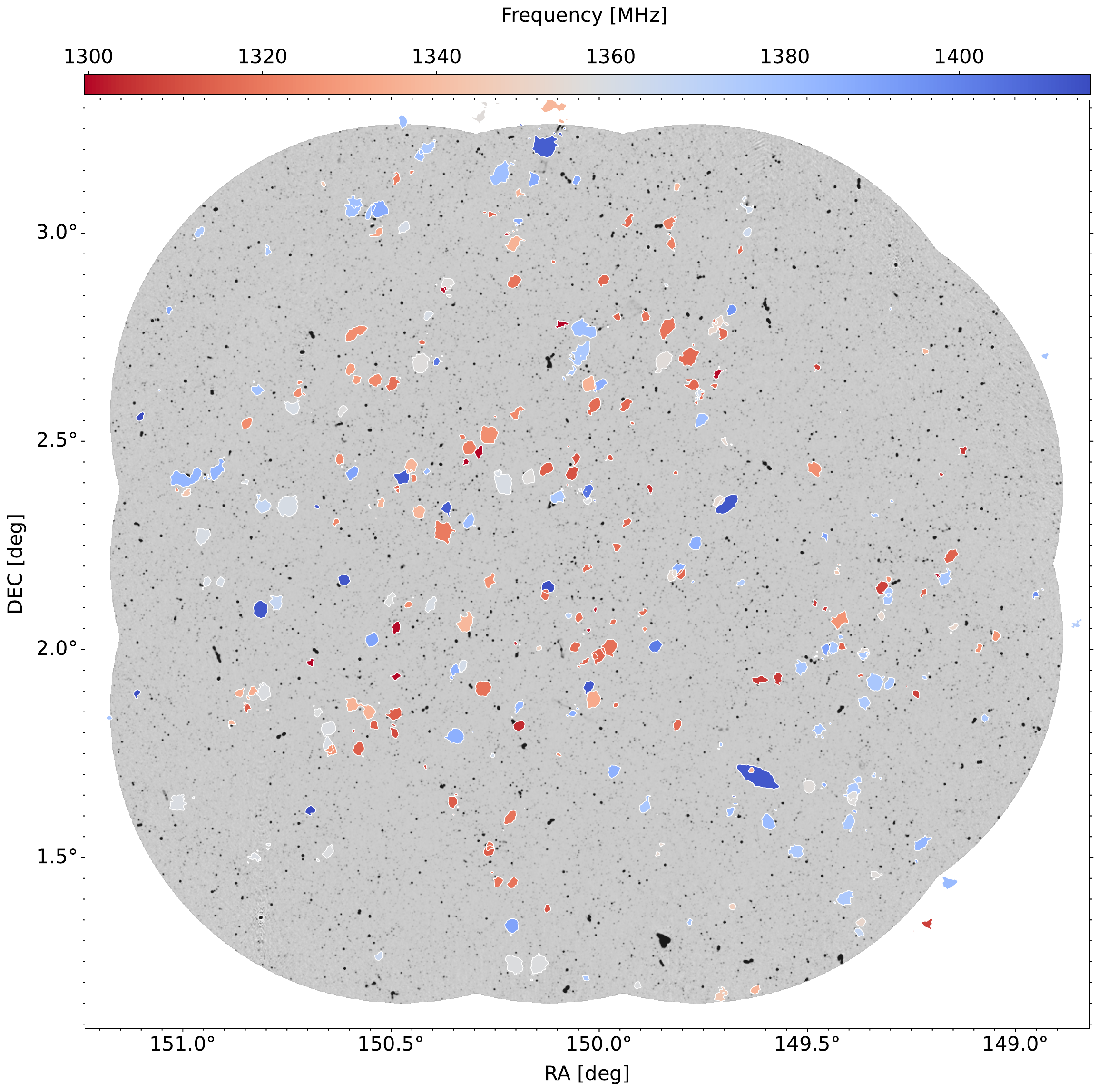}
    \caption{Contour islands of the HI distribution found by LESHI in the COSMOS field for a sample of 293 galaxies, colour coded by detection frequency, overplotted over radio continuum image. The contours are scaled up by a factor of two for better visibility and do not represent real sizes.}
\label{fig:cat_map}
\end{figure*}

\section{Examples of detections}

\begin{figure*}
    \centering
    \includegraphics[width=\linewidth]{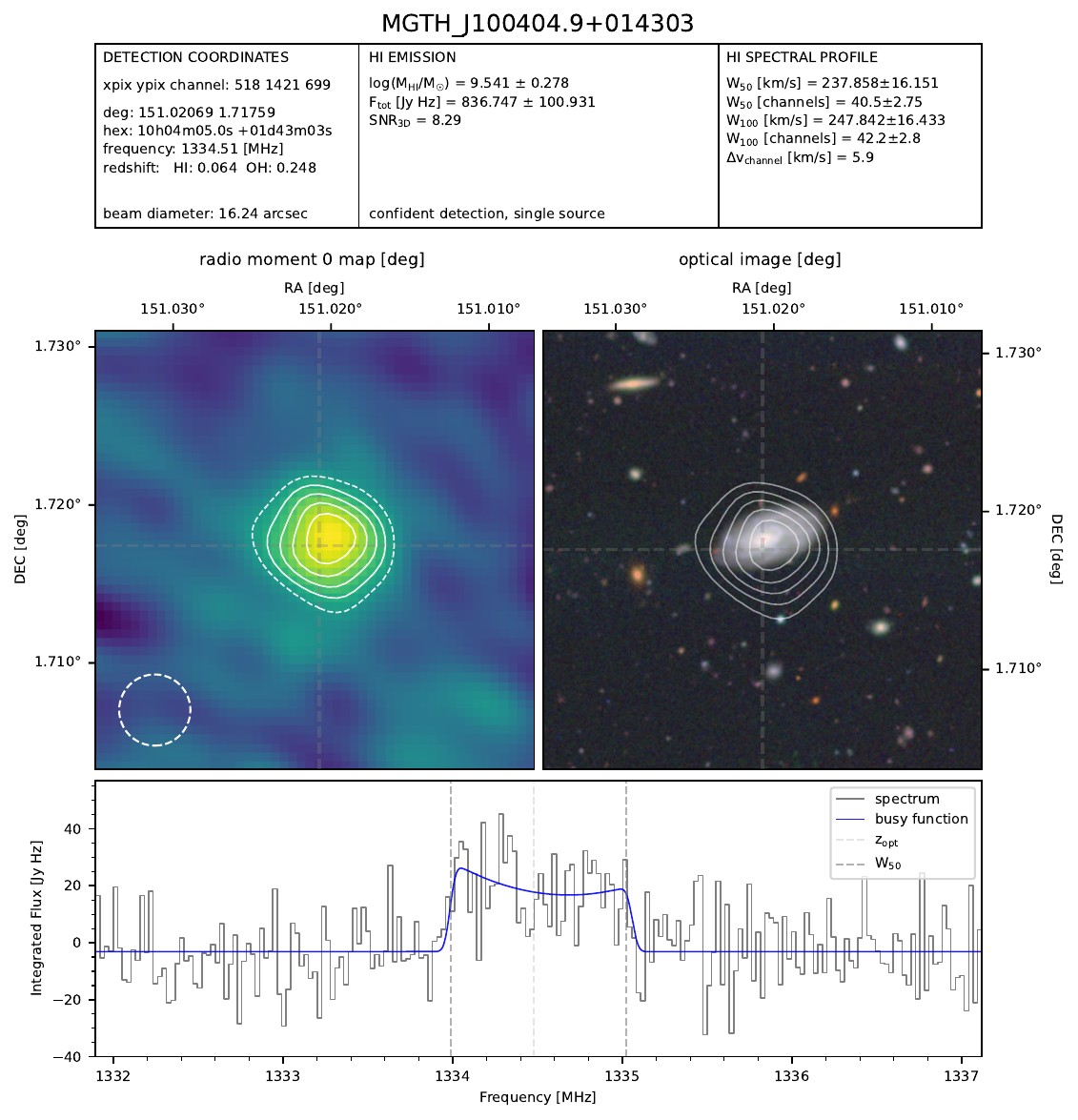}
    \caption{The only detection that was found by SoFiA and missed by LESHI.}
\label{fig:only_sofia}
\end{figure*}

\begin{figure*}
    \centering
    \includegraphics[width=\linewidth]{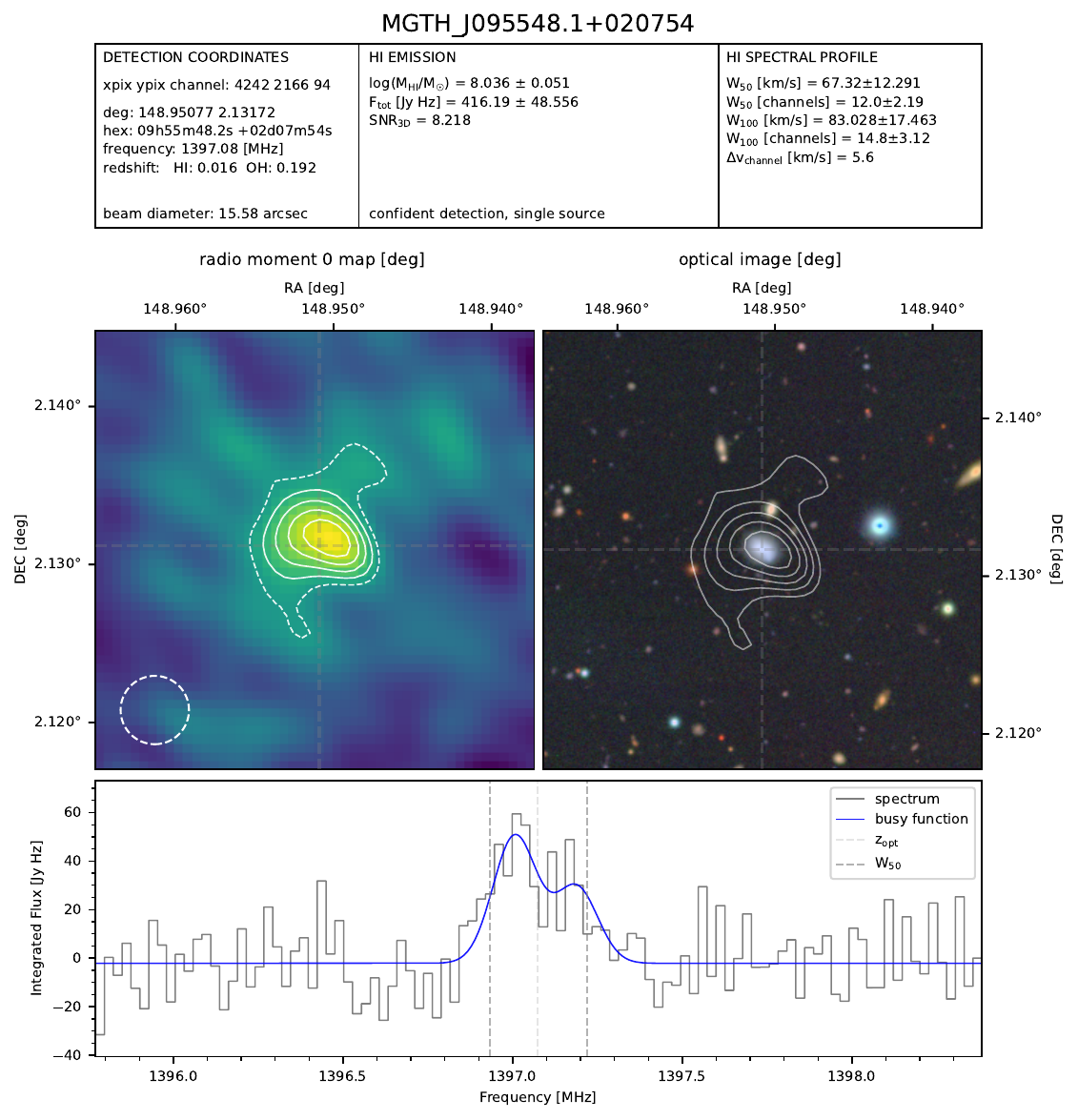}
    \caption{Example of a detection that was found by LESHI, but not by any other source finder.}
\label{fig:only_leshi_det}
\end{figure*}

\begin{figure*}
    \centering
    \includegraphics[width=\linewidth]{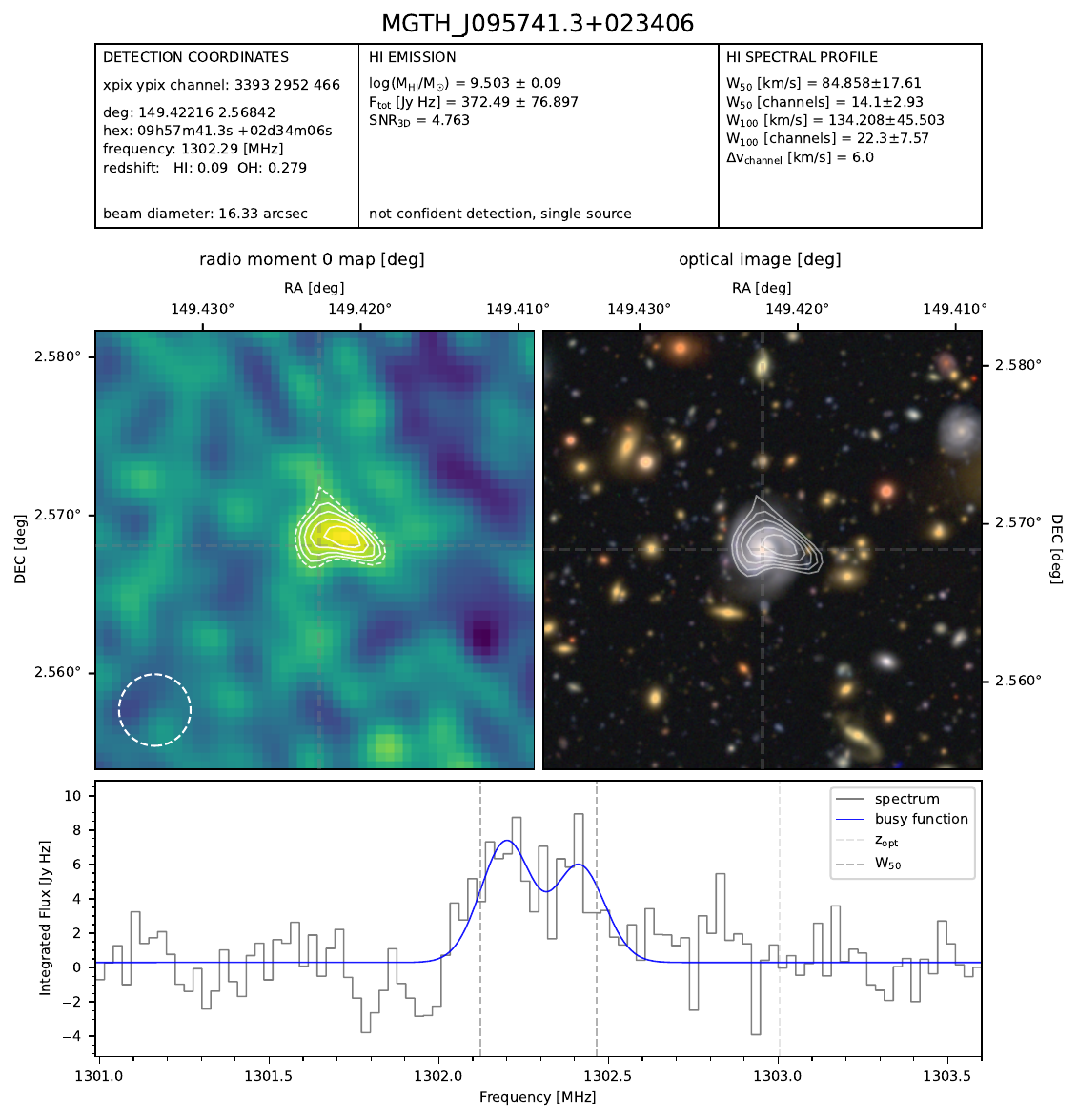}
    \caption{Example of a detection with the "low confidence" flag. The \HI\ detection has an optical counterpart, however the detection has low signal-to-noise and its distribution does not completely follow the optical data.}
\label{fig:low_confidence}
\end{figure*}

\begin{figure*}
    \centering
    \includegraphics[width=\linewidth]{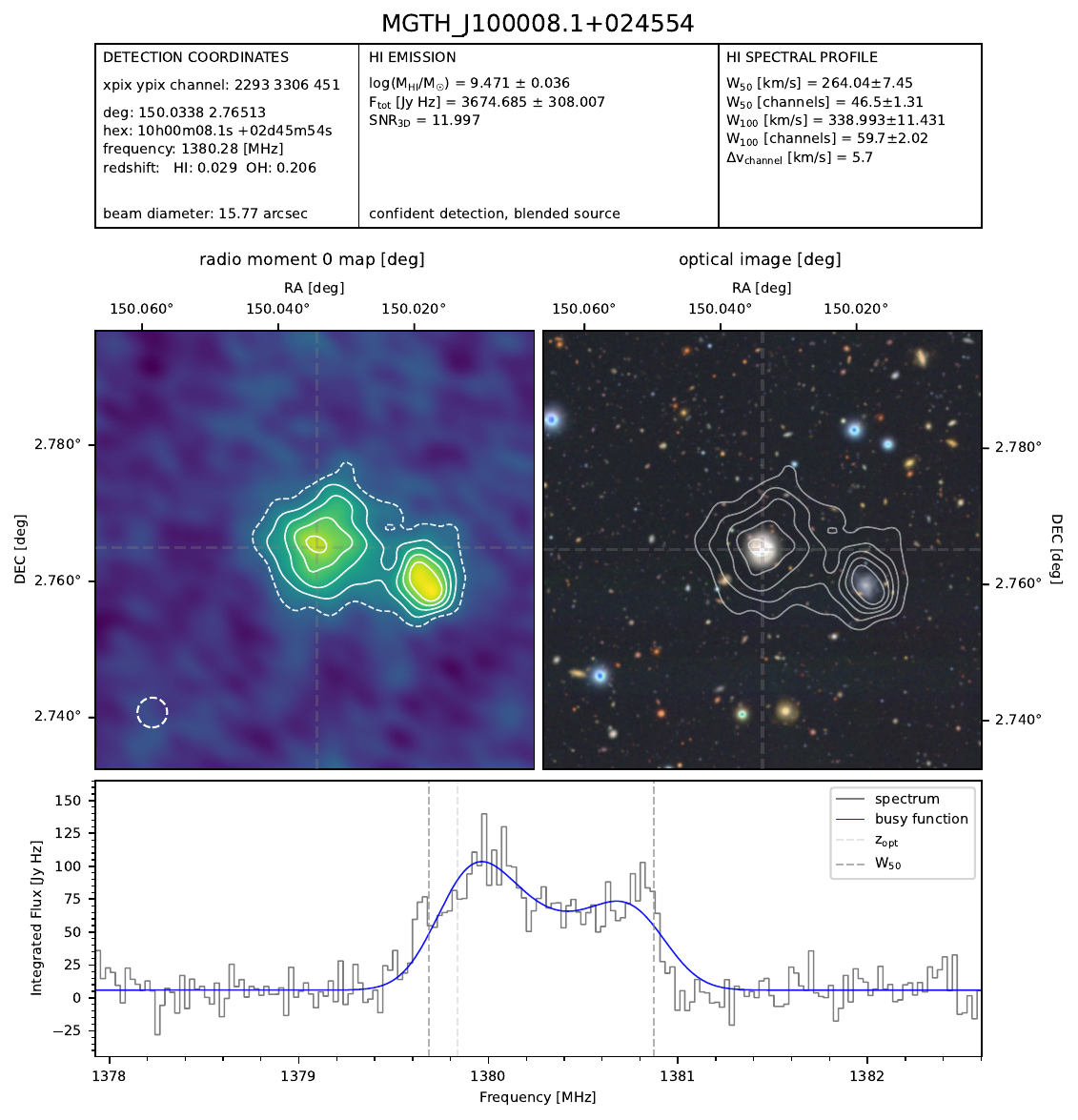}
    \caption{Example of a detection with the "blended source" flag. The detection contains emission belonging to two galaxies which are interacting and their \HI\ contents cannot be discerned.}
\label{fig:blended}
\end{figure*}

\begin{figure*}
    \centering
    \includegraphics[width=\linewidth]{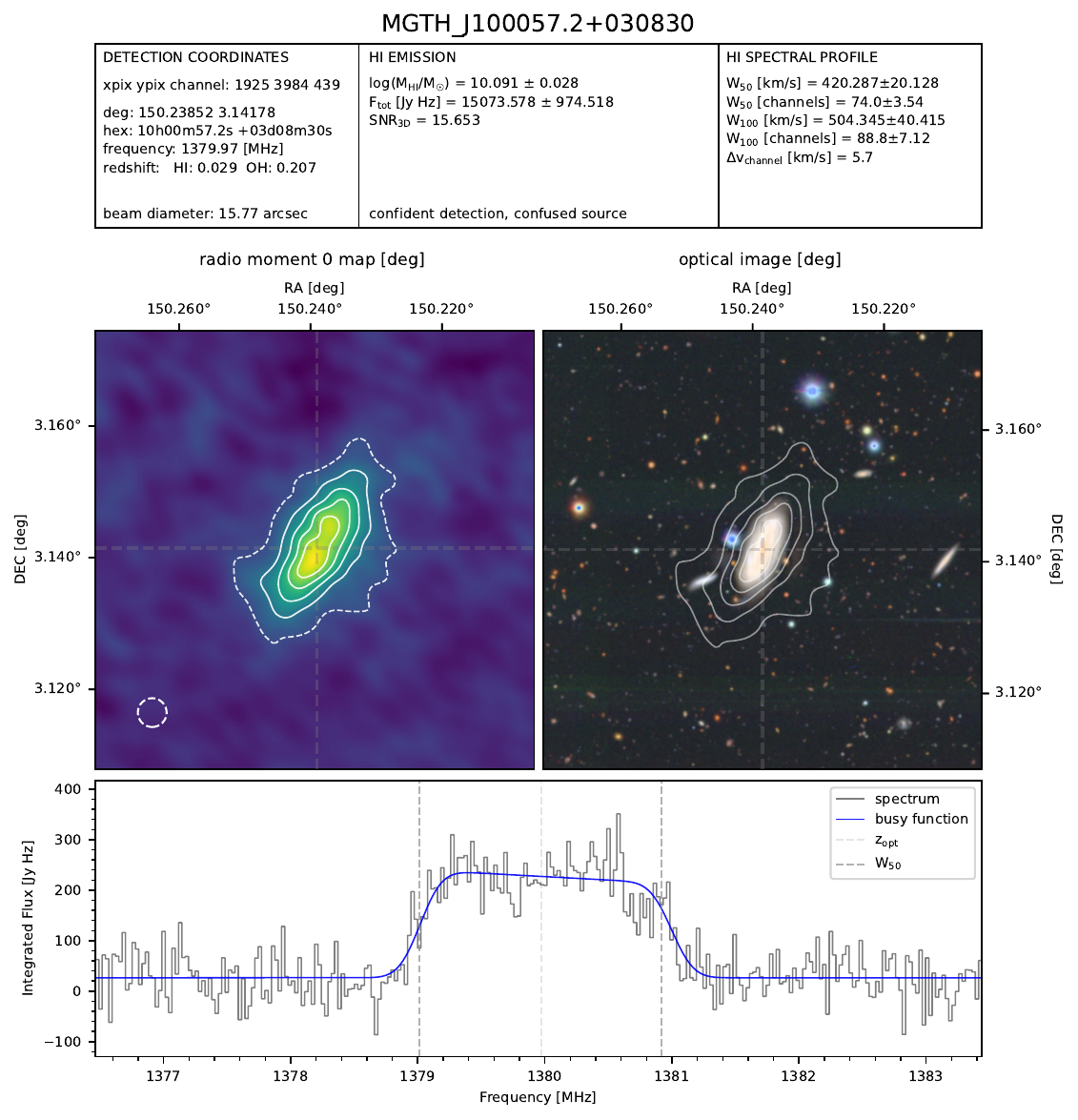}
    \caption{Example of a detection with the "confused source" flag. Within the distribution of the detected emission there is more than one galaxy with similar redshift in the optical data, however it is unclear from the \HI\ emission whether it is a blended source or overlapped projection.}
\label{fig:confused}
\end{figure*}

\section{Comparison with Early Science Data}
\begin{figure}
    \centering
    \includegraphics[width=\linewidth]{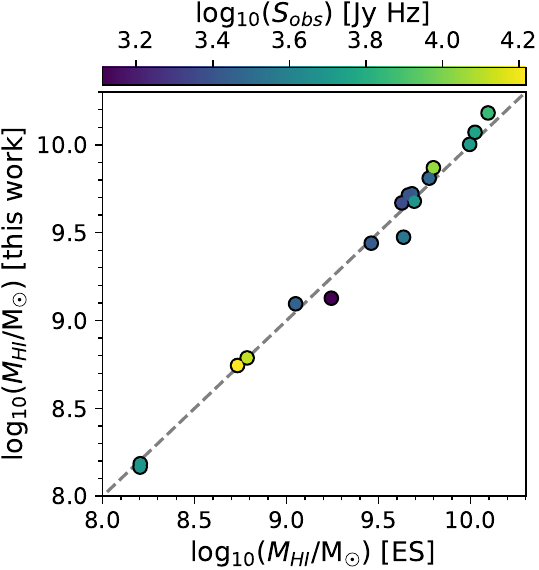}
    \caption{\HI\ mass measured in this work plotted against the \HI\ mass measured in \citet{2021MNRAS.508.1195P}, colour coded by the observed flux.} 
\label{fig:es}
\end{figure}

\newpage
\newpage
\bsp	
\label{lastpage}
\end{document}